\documentclass[usenatbib]{mn2e}
\usepackage[latin1]{inputenc}
\usepackage[T1]{fontenc}
\usepackage{graphicx}
\usepackage{natbib}
\usepackage[sumlimits, nointlimits, namelimits]{amsmath}
\usepackage{url}
\usepackage{rotating}

\renewcommand{\thefootnote}{\fnsymbol{footnote}}

\title[Satellite galaxies in hydrodynamical simulations] {Satellite
  galaxies in hydrodynamical simulations of Milky Way sized galaxies}
\author[M. Wadepuhl \& V. Springel] {Markus Wadepuhl\footnotemark[1]$^1$ and Volker
  Springel$^{1,2,3}$\\ $^1$Max-Planck-Institut for Astrophysics,
  Karl-Schwarzschild-Strasse 1, 85740 Garching bei M\"{u}nchen, Germany\\
$^2$Heidelberg Institute for Theoretical Studies,
  Schloss-Wolfsbrunnenweg 35, 69118 Heidelberg, Germany\\ 
$^3$Zentrum f\"{u}r Astronomie der Universit\"{a}t Heidelberg,
    M\"{o}nchhofstr. 12-14, 69120 Heidelberg, Germany
}

\begin{document}

\maketitle
\newcommand{\setthebls}{
}

\setthebls

\begin{abstract}
Collisionless simulations of the CDM cosmology predict a plethora of
dark matter substructures in the halos of Milky Way sized galaxies, yet
the number of known luminous satellites galaxies is very much smaller, a
discrepancy that has become known as the `missing satellite problem'.
The most massive substructures have been shown to be plausibly the hosts
of the brightest satellites, but it remains unclear which processes
prevent star formation in the many other, purely dark substructures. We
use high-resolution hydrodynamic simulations of the formation of Milky
Way sized galaxies in order to test how well such self-consistent models
of structure formation match the observed properties of the Galaxy's
satellite population. For the first time, we include in such
calculations feedback from cosmic rays injected into the star forming
gas by supernovae as well as the energy input from supermassive black
holes growing at the Milky Way's centre and its progenitor systems.  We
find that non-thermal particle populations quite strongly suppress the
star formation efficiency of the smallest galaxies. In fact, our cosmic
ray model is able to reproduce the observed faint-end of the satellite
luminosity function, while models that include only the effects of
cosmic reionization, or galactic winds, do significantly worse. Our
simulated satellite population approximately matches available kinematic
data on the satellites and their observed spatial distribution.  We
conclude that a proper resolution of the missing satellite problem
likely requires the inclusion of non-standard physics for regulating
star formation in the smallest halos, and that cosmic reionization alone
may not be sufficient.
\end{abstract}

\section{Introduction}

\renewcommand{\thefootnote}{\fnsymbol{footnote}}
\footnotetext[1]{E-mail: wadepuhl@mpa-garching.mpg.de}

The leading $\Lambda$CDM cosmology predicts that galaxies form
hierarchically in a `bottom up' fashion
\citep[e.g.][]{WhiteRees1978,GalaxyFormation}, where small perturbations
in the dark matter density distribution collapse earlier than larger
perturbations, and low-mass halos grow by smooth accretion or mergers
with other halos, successively building up ever bigger structures. But
structures falling into bigger systems during this process are not
always disrupted completely. As N-body simulations show, the inner cores
of infalling objects often survive the various disruptive effects acting
on them, like tidal truncation, tidal shocking or ram-pressure
stripping. It is believed that the observed dwarf galaxies orbiting
around the Milky Way (MW) are examples of such surviving remnants.

Based on the first generation of very high resolution collisionless CDM
simulations, \citet{FirstMSP} and \citet{Moore1999} pointed out a very
striking apparent discrepancy between theoretical predictions for such
satellite systems and actual observations. Given the very large number
of predicted dark matter substructures, there appears to be a dearth of
luminous satellites in the Milky Way. In fact, the cumulative number of
observed satellite galaxies and of predicted substructures above a given
circular velocity value differed by a factor of $\sim 10$.  This has
become known as the `missing satellite problem'. 

The initial analysis of \citet{Moore1999} and \citet{FirstMSP} may have
overstated the magnitude of the discrepancy, both because of
uncertainties in assigning correct circular velocity values to the
observed satellites \citep{Stoehr2002} and because a number of
additional faint satellites have been discovered meanwhile in the MW
\citep[see for example][]{SatData9, SatData11, SatData13, SatKinematics, SatData14,
SatData15, SatData16, SatData17, SatData18, SatData19}.
However, there is a consensus that the many low-mass
satellites predicted by the N-body simulations need to be strongly
suppressed in luminosity, otherwise a significant discrepancy with the
observed satellite luminosity functions results, that, if confirmed, may
in principle even be used to rule out cold dark matter.

With increasing numerical resolution, the missing satellite problem has
become more acute. Modern cosmological dark matter simulations of Milky
Way sized halos \citep{Diemand2008,Aquarius,Stadel2009} resolve up to
$\sim 300,000$ dark matter substructures, while the number of observed
satellite galaxies still comprises just a few dozens. We note that the
modern dark matter only simulations are even able to resolve
substructures inside substructures, and interestingly, there is also some
observational evidence for a satellite possibly orbiting around another
satellite \citep{SatData16}.

\citet{FirstMSP} did not only raise the missing satellite problem, they
were also among the first to suggest a potential solution to this
issue. In particular, they proposed that star formation inside low mass
halos could be suppressed because of photo evaporation of gas due to a
strong intergalactic ionizing UV background. This would keep most of the
orbiting satellites dark and render them visually unobservable. Indeed,
the simple filtering mass model of \citet{Gnedin2000} for the impact of
a UV background on the cooling efficiency of small halos predicts a
quite sizable effect, with a nearly complete suppression of cooling in
all halos with circular velocity below $50\,{\rm km\, s^{-1}}$. However,
recent full hydrodynamical simulations have not confirmed this
\citep{Hoeft2006,Okamoto2008}. They find a considerably weaker
effect, where only halos with circular velocities less than $\sim
25\,{\rm km\, s^{-1}}$ are affected. This also casts some doubt about
the faint-end results of numerous semi-analytic models for the satellite
population \citep[e.g.][]{Benson2002,Kravtsov2004}, which typically
employed the filtering mass formalism and hence assumed an overly strong
effect of the UV background. We will reexamine this question in this
work based on our cosmological hydrodynamic simulations of Milky Way
formation, which include a treatment of cosmic reionization.

Another possible solution for the satellite problem was proposed by
\citet{Redef_MSP} who suggested that not the satellite mass at the
present epoch determines whether a satellite would be luminous or not,
but rather the maximum mass it had before accretion onto the Milky Way's
halo. This is based on the idea that tidal stripping and ram pressure
unbinds the gas from an infalling satellite and thus stalls any further
star formation. With this assumption, the stellar mass at the time of
accretion is essentially retained until the present epoch, and it
becomes a question of allowing high-redshift star formation only in
satellites above a sufficiently high mass threshold.

A more radical conjecture is that the properties of the dark matter
particles may have to be changed. Instead of having negligible velocity
dispersion at the time of decoupling, we may instead be dealing with
(slightly)  warm dark matter (WDM). This can suppress the abundance
of low mass structure considerably \citep[e.g.][]{Colin2000}, but
provided the particle mass is not lower than $\sim 1\,{\rm keV}$
a sufficiently large number of substructures still survives to
explain the observed satellite abundance \citep{MacioFontanot2010}.

In the most recent works on the subject, a number of interesting and
encouraging results have been obtained.  Observationally, it has been
discovered that the satellites all have approximately the same central
mass density (within 300 to 600 pc), independent of their luminosity
\citep{Gilmore2007,Strigari2008}. Explaining this central density
threshold has become an important additional challenge for theoretical
models. Also, a significant number of new faint satellites have been
discovered with the help of the SDSS \citep{SatData9, SatData11, SatData13, SatKinematics, SatData14,
SatData15, SatData16, SatData17, SatData18, SatData19}, improving our
knowledge about the full satellite population significantly, but at the
same time also raising the question whether we may perhaps still be
missing large numbers of satellites at ultra low surface brightnesses.

On the theoretical side, refined treatments of the effects of
reionization, often coupled to the results of high-resolution
collisionless simulations have been used to model the satellite
population.  \citet{Maccio10} employed a number of different
semi-analytic models and low-resolution hydrodynamic simulations to
study the satellite luminosity function. Despite just invoking
photoheating as primary feedback process, they achieved reasonable
agreement for some of their models, leading them to argue that the
satellite problem may be solved. Similarly, \citet{Li2010} invoked a
strong impact of reionization in a semi-analytic model similar to those
applied to the Millennium Simulation \citep{Croton2006} to
reproduce the luminosity function of galaxies around the Milky
Way. \citet{Busha2010} used simple prescriptions for the impact of
inhomogeneous reionization on the satellite population, pointing out
that subtle changes in the assumptions about how reionization affects
star formation in small galaxies can lead to large changes in the
predicted number of satellites.

Recently, \citet{Strigari2010} examined the kinematics of five
well-measured Milky Way satellite galaxies and compared them to dark
matter satellites of the high-resolution simulations of the Aquarius
Project \citep{Aquarius}. They showed that these systems are fully
consistent with $\Lambda$CDM expectations and may be hosted in cored
dark matter structures with maximum circular velocities in the range 10
to $30\,{\rm km\,s^{-1}}$.  Interestingly, \citet{Bullock2009} pointed
out that the number of real satellite systems may in fact be much larger
than commonly believed, with the majority of them being so far
undetected because of their low surface brightness. In this scenario,
the `common mass scale' inferred for the observed satellites may in fact
just arise from a selection bias.

The first high-resolution hydrodynamic simulation able to directly
resolve the satellite population has recently been presented by
\citet{Okamoto2009}. They argue that the common mass scale identified in
the observations arises from early reionization at redshift around
$z\sim 12$, and that satellites that have not yet grown to a maximum
circular velocity of $\sim 12\,{\rm km\, s^{-1}}$ {\em by the time of
  reionization}, will not be able to make any stars later on. Even if
they grow above this threshold, \citet{Okamoto2009} predict them to
remain dark.

Despite all of this progress, it is evident that there remain many open
questions concerning the population of faint and ultra-faint satellite
galaxies orbiting around Milky Way like galaxies. Especially the
influence of different baryonic feedback processes on the luminosity
function of the simulated satellites has not been investigated in
sufficient detail. It is therefore far from clear whether photoheating
from a UV background and ordinary supernovae feedback are indeed the
correct physical solutions to the missing satellite problem. In fact, the
in part contradictory results that have been obtained with analytic
recipes to describe the impact of reionization suggest that more
accurate methodologies are required to reliably settle the issue.

We have therefore embarked on a research program where we use
high-resolution hydrodynamic simulations of the formation of Milky
Way-sized halos to shed more light on these questions, in particular by
investigating a variety of feedback processes known to be important in
galaxy formation. Besides the impact of reionization, these include
galactic winds and outflows, energy input by growing supermassive black
holes, or the non-thermal support of gas by cosmic rays or magnetic
fields. Ultimately we aim to reach similar numerical resolution as has
been obtained for recent collisionless simulations, even though this
goal may still be several years away.

In this work, we present some of our first results. We use several
well resolved hydrodynamical simulations of the formation of a Milky
Way sized galaxy to investigate the properties of the predicted
population of satellite galaxies, for different choices of the included
physics. Besides a default reference model that includes only a
treatment of radiative cooling, star formation, and cosmic reionization,
we consider also models that add galactic winds, supermassive black hole
growth, or cosmic ray injection by supernovae shock waves. By comparing
the simulation results with a comprehensive catalogue of the known Milky
Way satellites, we seek to determine which of these processes is most
important in shaping the satellite population.

This paper is organized as follows. In Section~\ref{sec:methodology}, we
describe the methodological details of our simulations, while the
observational knowledge about the satellites is briefly summarized in
Section~\ref{sec:observations}. Sections~\ref{sec:abundance},
\ref{sec:history} and \ref{sec:scaling} present the results for our
simulated populations of satellite galaxies, both with respect to
individual satellite histories as well as with respect to their population
as a whole. Our conclusions are summarized in
Section~\ref{sec:conclusions}.

\section{Methodology}\label{sec:methodology}

Our simulations are based on initial conditions originally constructed
for the Aquarius Project \citep{Aquarius} of the Virgo
Consortium. This project carried out highly resolved dark matter only
simulations of 6 different Milky Way-sized halos, at a variety of
different numerical resolutions. In the nomenclature of
\citet{Aquarius}, the halo and resolution level investigated here is
called `Aq-C-4', and has about 5.4 million dark matter particles in the
final virial radius.  The same object has also been studied in the
hydrodynamic simulations of \citet{Scannapieco2009}, albeit at the
considerably lower resolution (by a factor 8 in particle number)
corresponding to `Aq-C-5'. There it was found that this `C'-halo
produced the lowest bulge-to-disk ratio among the 6 candidate halos
selected in the Aquarius Project, making it a particularly good
candidate for the formation of a large disk galaxy. We note, however,
that we do not expect our choice of target halo to influence our
principal conclusions for the satellite population.

In this paper, we model the gas component with smoothed particle
hydrodynamics (SPH) and introduce the gas particles into the initial
conditions by splitting each original particle into a dark matter and
gas particle pair, displaced slightly with respect to each other (at
fixed center-of-mass) to arrive at a regular distribution of the mean
particle separations, and with a mass ratio corresponding to a baryon
fraction of 16 per cent.  The cosmological parameters are $\Omega_m =
0.25$, $\Omega_{\Lambda}=0.75$, $\sigma_8=0.9$ and $h=0.73$, the same
ones used as in the original Aquarius simulations, which are consistent
with the WMAP1 cosmological constraints.  A periodic box of size
$100\,h^{-1}{\rm Mpc}$ on a side is simulated, with varying spatial
resolution that `zooms in' on the formation of a single galaxy. In the
high-resolution region, we reach a mass simulation of $\approx 2\times
10^5 \,h^{-1}{\rm M_{\odot}}$ and $\approx 2\times 10^4 \,h^{-1}{\rm
  M_{\odot}}$ for dark matter and gas particles, respectively. A
constant comoving gravitational softening length of $\epsilon = 0.25 \,
h^{-1} \mathrm{kpc}$ was used for all high-resolution particles.

We employed the parallel TreeSPH code \textsc{gadget-3} for our runs,
which is an improved and extended version of \textsc{gadget-2}
\citep{Gadget2}. \textsc{gadget} calculates the long-range gravitational
field in Fourier space, and the short range forces in real space with a
hierarchical multipole expansion, based on a tree. This approach
guarantees a homogeneously high spatial resolution in the gravitational
force calculation and can be efficiently combined with an individual
timestep integration scheme. For the hydrodynamics, \textsc{gadget} uses
the `entropy formulation' of SPH \citep{Springel2002}, which is derived
from a variational principle and simultaneously conserves energy and
entropy where appropriate.

In the hydrodynamic part of \textsc{gadget}, different physical
processes besides ordinary gas dynamics are calculated. Most
importantly, these are radiative cooling, star formation and its
regulation by supernovae feedback processes \citep{SFR_paper}.  The code
can optionally also model black hole growth and \citep{BH_paper} and
cosmic ray physics \citep{CR_Paper}. We shall now briefly describe the
physics modules we used.

\subsection{Star formation model}

Radiative cooling is followed for a primordial mixture of helium and
hydrogen under the assumption of collisional ionization equilibrium,
using a formulation as in \citet{Katz1996}.  A spatially uniform,
ionizing UV background is introduced with an amplitude and time
evolution described by an updated version of \citet{Haardt1996}, leading
to reionization of the universe by redshift $z\simeq 6$.

To model star formation, we use the hybrid multiphase model for star
formation and supernova feedback introduced by \citet{SFR_paper}, in
which every sufficiently dense gas particle is treated as a
representative region for the multiphase structure of the interstellar
medium (ISM). These hybrid particles are pictured to be comprised of
cold dense clouds in pressure equilibrium with a hot ambient gas phase,
where only the clouds contribute material available for star
formation. Mass and energy exchange processes between these two phases
are computed by simple differential equations, as described in
\citet{SFR_paper}, giving rise to an effective equation of state that
regulates the dense gas of the ISM. Collisionless star particles are
spawned stochastically from this star-forming phase according to a local
estimate of the star formation rate. The gas consumption timescale of
the model is calibrated such that it reproduces the Kennicutt law \citep{Kennicutt}
between star formation rate and gas surface density
observed in low-redshift disk galaxies.

\subsection{Wind model}

Even though the above ISM model reproduces the star formation rates
observed in disk galaxies, it overproduces the total amount of stars
formed when applied in cosmological simulations. This is likely related
to its inability to reproduce the observed galaxy-scale outflows seen
around many starbursting galaxies. We have therefore also investigated
the phenomenological model for galactic winds introduced by
\citet{SFR_paper}. We adopted a constant wind velocity of $v_{\rm wind}=
484\,{\rm km\,s^{-1}}$ and mass loading factor of $\eta=2$, i.e.~the
mass flux of the wind is twice the star formation rate.  Individual gas
particles were stochastically added to the galactic wind by changing
their velocity to the prescribed wind velocity.  We adopted an
anisotropic distribution for the wind direction, launching it
preferentially perpendicular to the disk.

The wind model causes an outflow of gas from the dense gaseous disc,
transporting energy, matter and heavy elements out of the disc in
proportion to the star formation rate. Not only the central galaxy is
affected by the winds, all the star forming satellite galaxies produce
winds, which will more quickly deplete their gas content. The effect of
galactic winds is actually expected to be more effective in these low
mass systems as their potential well is much shallower than the one of
the main galaxy \citep{Dekel1986}, which should increase the mass loss
due to outflows. Including winds in the simulation thus appears in
principle promising as a mechanism to reduce the number of luminous
satellites, but the effect may sensitively depend on how the wind
properties are scaled with galaxy size \citep[e.g.][]{Oppenheimer2006}.

\subsection{Black hole model}

Supermassive black holes are thought to reside at the centres of most if
not all spheroidal galaxies. The tight relation between their masses and
the velocity dispersion of their hosting galaxies suggest a tight
evolutionary link, which is probably established by a self-limited
growth mechanism in which the energy output of a growing black hole
eventually terminates its further growth and the surrounding star
formation, for example by expelling gas from the central region of the
galaxy. Hydrodynamical simulation models have been successfully used to
model this process in detail \citep{DiMatteo2005,BH_paper}, and led to
quite successful unified models of the formation of spheroidal galaxies
\citep{Hopkins2006}.

While it is unclear whether the influence of the Milky Way's black hole
has affected the formation of other components of the Galaxy besides the
central bulge, it appears possible that the heating effects from
different quasar episodes during the growth history of the MW's
supermassive black hole have had an impact on the satellite population
as well, for example by heating the environment of progenitor halos
through strong quasar driven outflows. Indeed, in theoretical models
outflow feedback has been found to be violent enough to be able to
strongly affect even neighbouring galaxies
\citep{Scannapieco2001,Thacker2002}.

To study such effects, we have adopted the techniques introduced by
\citet{BH_paper} for tracking black hole growth and its associated
energy feedback in cosmological simulations. In brief, we periodically
call a FoF group finding algorithm that identifies newly formed halos
that do not contain a black hole yet. If such a halo is sufficiently
massive, its densest gas particle is converted to a black hole seed of
mass $M_{\rm BH}=10^5\,h^{-1}{\rm M}_\odot$. The black hole particles
are treated as sinks particles that accrete gas from their surroundings
at a rate estimate with a simple Bondi-Hoyle prescription, limited to
the Eddington rate. The black hole accretion is assumed to have a fixed
radiative efficiency of $0.1\, \dot  M c^2$, and 5\% of the
produced radiation are assumed to couple thermally to the gas
surrounding the black hole. This energy feedback can eventually drive a
quasar-driven outflow and regulates the mass growth of the black
holes. We also allow for two black holes to merge with each other once
they get sufficiently close to each other.

\subsection{Cosmic ray model}

In the interstellar medium of our own Galaxy, it is believed that
thermal pressure, cosmic rays and magnetic fields are roughly in
equipartition. Even though it is hence known that non-thermal particle
populations play an important role in regulating the gas dynamics of the
ISM, this component has usually been ignored in studies of galaxy
formation. The cosmic ray particles may originate in acceleration
processes in high Mach number shocks in supernova remnants or could be
produced in structure formation shock waves. We here focus on cosmic ray
injection associated with supernovae, and consider only Coulomb and
hadronic interactions as loss processes for the cosmic ray particles
\citep{Ensslin2007}.

A numerical treatment of the cosmic ray component is rather complicated
as in principle their full, in general anisotropic distribution function
has to be modeled. Also, the motion of the cosmic ray fluid is tightly
coupled to the magnetic field, which in turn is non-trivial to
calculate accurately. We therefore employ the subresolution model described
by \cite{CR_Paper}, which has already been successfully employed in
previous work \citep{Pfrommer2007, Pfrommer2008a, Pfrommer2008b}.

\begin{table*}
	\centering
	\small{
	\begin{tabular}{llll} 
		Label  & Mass resolution (gas) & Gravitational softening & Physics\\
		\hline \hline
		Ref    & $5.14 \times 10^4 \;\mathrm{M_\odot}$ & $0.34 \;\mathrm{kpc}$ & star formation, supernova feedback$^{(a)}$ \\
		BH     & $5.14 \times 10^4 \;\mathrm{M_\odot}$ & $0.34 \;\mathrm{kpc}$ & star formation, supernova feedback$^{(a)}$, AGN feedback$^{(b)}$ \\
		Wind   & $5.14 \times 10^4 \;\mathrm{M_\odot}$ & $0.34 \;\mathrm{kpc}$ & star formation, supernova feedback$^{(a)}$, galactic winds$^{(a)}$ \\
		CR     & $5.14 \times 10^4 \;\mathrm{M_\odot}$ & $0.34 \;\mathrm{kpc}$ & star formation, supernova feedback$^{(a)}$, AGN feedback$^{(b)}$, cosmic rays$^{(c)}$ \\
		LowRes & $4.11 \times 10^5 \;\mathrm{M_\odot}$ & $0.68 \;\mathrm{kpc}$ & star formation, supernova feedback$^{(a)}$, AGN feedback$^{(b)}$\\
		\hline \hline
	\end{tabular}
	\caption{Overview of the simulations used in this work. The
          label of each simulation will be used throughout the rest of
          this paper. The different physics models we use are described
          in detail in (a) \citet{SFR_paper}, (b) \citet{BH_paper}, and
          (c) \citet{CR_Paper}.
	\label{tab:simulations} }}
\end{table*}

The basic assumption of our CR model \citep{Ensslin2007,CR_Paper} is
that the momentum spectrum of cosmic rays can be well represented by a
simple power law of the form
\begin{equation}
\frac{\text{d}^2N}{\text{d}p\;\text{d}V} = Cp^{-\alpha}\theta(p-q),
\end{equation}
where $C$ gives the normalization, $q$ is a low momentum cut-off, and
$\alpha$ is the power law slope. We assume that the cosmic rays are
dominated by protons, and that the local magnetic field is sufficiently
tangled to effectively lock in the CRs to the gas.
The pressure of the cosmic ray
population is then given by
\begin{equation}
P_{\text{CR}} = \frac{Cm_pC^2}{6} \mathcal B_{\frac{1}{1+q^2}}\left(\frac{\alpha-2}{2},\frac{3-\alpha}{2}\right),
\end{equation}
while the number density is simply $n_{\text{CR}}= Cq^{1-\alpha}/(\alpha-1)$. Here
\begin{equation}
\mathcal B_n(a,b) \equiv \int_0^n x^{a-1}(1-x)^{b-1} \text{d}x
\end{equation}
denotes incomplete Beta functions. Based on the findings of
\citet{CR_Paper}, we expect this additional pressure to be especially
influential in low mass satellites, because it here can substantially
reduce the density of the ISM.  This should effectively reduce the
number of low luminosity satellites.

We model the decay of the cosmic ray population by accounting for
coulomb cooling and catastrophic hadronic losses as described in
\citet{CR_Paper}. Note that the `cosmic ray cooling' mediated by these
effects can occur on a different timescale as the ordinary thermal
cooling. In particular, gas can end up being cosmic ray pressure
supported after having lost much of its thermal support through
radiative cooling.  Finally, we adopt a simple source function for the
injection of new cosmic ray particles, which we link directly to the
star formation rate. This is motivated by observations of supernova
remnants \citep{Aharonian2006}, where a large fraction of the
supernovae energy is seen to initially appear as cosmic rays.

\subsection{Simulation set and analysis}

In Table~\ref{tab:simulations}, we summarize the primary properties of
the simulations we analyze in this work. We consider four different
high-resolution simulations of the same initial conditions,
corresponding to the Aq-C-4 halo, but carried out with different physics
in the baryonic sector. Our reference calculation (labeled
`\textsc{Ref}') includes star formation and supernova feedback as
described by our multi-phase model, as well as ordinary radiative
cooling and heating by a UV background that reionizes the universe by
redshift $z=6$. We have repeated this calculation by adding in turn each
of the three additional feedback models described above. This yields the
three simulations `\textsc{BH}' (with black hole growth and feedback),
'\textsc{Wind}' (with the phenomenological wind model), and
'\textsc{CR}' (with cosmic ray physics). Our primary simulation set is
composed of these four simulations. They are of equal numerical
resolution and hence allow a relatively clean assessment of the impact
of the different physics components on the satellite population. We note
that our primary aim in this work is not to construct a best-fitting
model for the Milky Way, as this may require a combination of the
different physics models and a fine-tuning of their free
parameters. Rather we want to highlight the importance of different
physics for the satellite population.

We also briefly consider a further simulation, labeled
`\textsc{LowRes}'. This is a rerun of our reference model at lower
resolution, corresponding to `Aq-A-5'. We use this simulation for an
assessment of the numerical resolution and convergence limits of our
simulations.

\begin{table*}
	\centering
	\small{
	\begin{tabular}{lllllll} 
		Label  & \begin{sideways}$\alpha_{2000}$\end{sideways} & \begin{sideways}$\delta_{2000}$\end{sideways} & \begin{sideways}D (kpc)\end{sideways} & \begin{sideways}M$_{\mathrm{V,tot}}$ (mag)\end{sideways} & \begin{sideways}$\mu_{0,V}$ (mag/arcsec$^2$)\end{sideways} & \begin{sideways}Mass ($10^{6} \mathrm{M_\odot}$)\end{sideways}\\
		\hline \hline
		BooI$^{(\mathrm{e})}$    & $14^{\mathrm{h}}00^{\mathrm{m}}$ & $+14^{\circ}30'$ & $66\pm3$ & $-6.3\pm0.2$ & $27.5\pm0.3$       & --\\
		BooII$^{(\mathrm{e})}$   & $13\;\,58$ & $+12\;51$ & $42\pm8$             & $-2.7\pm0.9$         & $28.1\pm1.6$                & --\\
		Carina$^{(\mathrm{b})}$  & $60\;\,42$ & $-50\;58$ & $101\pm5$            & $-9.3$               & $25.5\pm0.4^{(\mathrm{h})}$ & $13$\\
		Com$^{(\mathrm{e})}$     & $12\;\,27$ & $+23\;54$ & $44\pm4$             & $-4.1\pm0.5$         & $27.3^{+0.7}_{-0.6}$        & $1.2\pm0.4^{(\mathrm{f})}$\\
		CVnI$^{(\mathrm{e})}$    & $13\;\,28$ & $+33\;33$ & $218\pm10$           & $-8.6^{+0.2}_{-0.1}$ & $27.1\pm0.2$                & $27\pm4^{(\mathrm{f})}$\\
		CVnII$^{(\mathrm{e})}$   & $12\;\,57$ & $+34\;19$ & $160^{+4}_{-5}$      & $-4.9\pm0.5$         & $26.1^{+0.7}_{-0.6}$        & $2.4\pm1.1^{(\mathrm{f})}$\\
		Draco$^{(\mathrm{e})}$   & $17\;\,20$ & $+57\;55$ & $76\pm5$             & $-8.8\pm0.2$         & $25.5\pm0.2$                & $22$\\
		Fornax$^{(\mathrm{b})}$  & $02\;\,40$ & $-34\;27$ & $138\pm8$            & $-13.2$              & $23\pm0.3^{(\mathrm{h})}$   & $68$\\
		Her$^{(\mathrm{e})}$     & $16\;\,31$ & $+12\;48$ & $132\pm12$           & $-6.6\pm0.3$         & $27.2^{+0.6}_{-0.5}$        & $7.1\pm2.6^{(\mathrm{f})}$\\
		LeoA$^{(\mathrm{b})}$    & $09\;\,59$ & $+30\;45$ & $690\pm100$          & $-11.4$              & --                          & $11$\\
		LeoI$^{(\mathrm{b})}$    & $10\;\,08$ & $+12\;19$ & $250\pm30$           & $-11.9$              & $22.4\pm0.3^{(\mathrm{h})}$ & $22$\\
		LeoII$^{(\mathrm{b})}$   & $11\;\,13$ & $+22\;09$ & $205\pm12$           & $-9.6$               & $24.0\pm0.3^{(\mathrm{h})}$ & $9.7$\\
		LeoIV$^{(\mathrm{e})}$   & $11\;\,33$ & $-00\;32$ & $160^{+15}_{-14}$    & $-5.0^{+0.6}_{-0.5}$ & $27.5^{+1.3}_{-1.2}$        & $1.4\pm1.5^{(\mathrm{f})}$\\
		LeoV$^{(\mathrm{j})}$    & $11\;\,31$ & $+02\;13$ & $180$                & $-4.3$               & $27.5\pm0.5$                & --\\              
		LeoT$^{(\mathrm{e})}$    & $09\;\,35$ & $+17\;03$ & $407\pm38$           & $-7.1^{(\mathrm{c})}$& $26.9^{(\mathrm{c})}$       & $8.2\pm3.6^{(\mathrm{f})}$\\
		LiuI$^{(\mathrm{d})}$    & $10\;\,00$ & $+57\;30$ & $83.2^{+9.3}_{-8.4}$ & $-4.15$              & $28.8$                      & --\\
		LiuII$^{(\mathrm{d})}$   & $13\;\,29$ & $+28\;41$ & $75.9^{+8.5}_{-7.6}$ & $-3.91$              & $29.2$                      & --\\
		LMC$^{(\mathrm{a})}$     & $05\;\,24$ & $-69\;50$ & $49$                 & $-18.5^{(\mathrm{i})}$              & $20.7\pm0.1^{(\mathrm{h})}$ & $10.000^{(\mathrm{g})}$\\
		NGC6822$^{(\mathrm{b})}$ & $19\;\,45$ & $-14\;48$ & $490\pm40$           & $-15.2$              & $21.4\pm0.2^{(\mathrm{h})}$ & $1640$\\
		Pegasus$^{(\mathrm{b})}$ & $23\;\,29$ & $+14\;45$ & $955\pm50$           & $-12.9$              & --                          & $58$\\
		Phoenix$^{(\mathrm{b})}$ & $01\;\,51$ & $-44\;27$ & $445\pm30$           & $-10.1$              & --                          & $33$\\
		PscI$^{(\mathrm{l})}$	 & $23\;\,19$ & $0\;0$	  & $80$				 & -- 					& --						  & $0.1$\\
		PscII$^{(\mathrm{m})}$	 & $22\;\,58$ & $5\;57$	  & $180$				 & $-5.0$				& --						  & --\\
		Sag$^{(\mathrm{b})}$     & $18\;\,55$ & $-30\;29$ & $24\pm2$             & $-13.4$              & $25.4\pm0.2^{(\mathrm{h})}$ & $150^{(\mathrm{i})}$\\
		Sculpor$^{(\mathrm{b})}$ & $01\;\,00$ & $-72\;50$ & $79\pm4$             & $-11.1$              & $23.7\pm0.4^{(\mathrm{h})}$ & $6.4$\\
		Seg1$^{(\mathrm{e})}$    & $10\;\,07$ & $+16\;04$ & $23\pm2$             & $-1.5^{+0.6}_{-0.8}$ & $27.6^{+1.0}_{-0.7}$        & --\\
		Seg2$^{(\mathrm{k})}$	 & $02\;\,19$ & $+20\;10$ & $35$				 & $-2.5\pm0.2$			& --						  & $0.55^{+1.1}_{-0.3}$\\
		Seg3$^{(\mathrm{m})}$    & $21\;\,21$ & $+19\;07$ & $16$				 & $-1.2$				& --						  & --\\
		Sextans$^{(\mathrm{b})}$ & $10\;\,13$ & $-01\;37$ & $86\pm4$             & $-9.5$               & $26.2\pm0.5^{(\mathrm{h})}$ & $19$\\
		SMC$^{(\mathrm{a})}$     & $00\;\,51$ & $-73\;10$ & $58$                 & $-17.1^{(\mathrm{i})}$              & $22.1\pm0.1^{(\mathrm{h})}$ & $400^{(\mathrm{g})}$\\
		Tucana$^{(\mathrm{b})}$  & $22\;\,42$ & $-64\;25$ & $880\pm40$           & $-9.6$               & $25.1\pm0.1^{(\mathrm{h})}$ & --\\
		UMaI$^{(\mathrm{e})}$    & $10\;\,35$ & $+51\;55$ & $96.8\pm4$           & $-5.5\pm0.3$         & $27.7^{+0.5}_{-0.4}$        & $15\pm4^{(\mathrm{f})}$\\
		UMaII$^{(\mathrm{e})}$   & $08\;\,07$ & $+63\;07$ & $30\pm5$             & $-4.2\pm0.5$         & $27.9\pm0.6$                & $4.9\pm2.2^{(\mathrm{f})}$\\
		UMi$^{(\mathrm{b})}$     & $15\;\,09$ & $+67\;13$ & $66\pm3$             & $-8.9$               & $25.5\pm0.5^{(\mathrm{h})}$ & $23$\\
		Wil1$^{(\mathrm{e})}$    & $10\;\,49$ & $+51\;03$ & $38\pm7$             & $-2.7\pm0.7$         & $26.1\pm0.9$                & --\\
		\hline \hline
	\end{tabular}
	\caption{Compilation of all presently known Milky Way satellite
          galaxies. The values are taken from the following studies: (a)
          \citet{SatData1}, (b) \citet{SatData2},(c)
          \citet{SatData9},(d) \citet{SatData11},(e) \citet{SatData13},
          (f) \citet{SatKinematics}, (g) \citet{Mass_Magellanic_Clouds},
          (h) \citet{SatData14}, (i) \citet{SatData15}, (j)
          \citet{SatData16}, (k) \citet{SatData17}, (l)
          \citet{SatData18}, (m) \citet{SatData19}. All errors are from
          the corresponding papers. The different columns list the
          position in galactic coordinates, the proper distance, the
          total V-band magnitude, the surface brightness and the total
          estimated mass of the individual satellites.}
	\label{tab:measured_satellites} }
\end{table*}

To analyze the time evolution of the simulated galaxies, several
snapshots were stored at different times. As a basic analysis step, the
snapshots were first processed by a group finding algorithm in order to
identify individual halos.  The group finding was done with a simple
friend-of-friends (FOF) algorithm \label{groupfinder} applied only to the dark
matter particles with a linking length equal to 20\% of their mean
particle spacing. The gas and star particles were linked to their
nearest dark matter particle. Next, each halo found in the first step
was subjected to a substructure detection procedure, for which we used
the {\small SUBFIND} algorithm \citep{Subfind} in a version extended to
allow a treatment of gas as well \citep{Dolag2009}. {\small SUBFIND}
calculates the local density everywhere and searches for substructure
candidates that are locally overdense. It then computes the
gravitational potential at the positions of all particles in the
candidate structures, and determines the subset of particles that are
gravitationally bound. In this way, only real physical structures are
found. To avoid noise from substructures composed of
only a few particles, only substructures containing at least $20$
particles were kept for further analysis. 

The gravitationally bound structures found by {\small SUBFIND} in this
way form our catalogue of simulated galaxies, including both central
galaxies as well as genuine satellites. For the simulated galaxies, we
applied a stellar population synthesis model \citep{Bruzual} to estimate
their luminosities and colours, based on the formation times and masses
of the star particles created in the simulations. We made no attempt to
account for the metallicity dependence of the stellar population
synthesis model or dust corrections.

\section{Observational knowledge}\label{sec:observations}

Before we present our simulation results, we summarize in this section
the most recent observational data with respect to the Milky Way's
satellite population. We will later use this comprehensive catalogue of
the known satellites together with predictions for their total number
over the whole sky when comparing with our simulations.

Table~\ref{tab:measured_satellites} gives a compilation of the
properties of all Milky Way satellites known today. The basic data of
the `classical' satellites, which were already known in 1998, are given
by \citet{SatData2} and were only slightly extended using
\citet{SatData15} who updated the data on the Small and the Large
Magellanic Cloud (SMC and LMC, respectively). Up to this time, the
number of known satellites was just $16$, but during the different data
releases\footnote{\url{http://www.sdss.org/dr6/index.html}}
\citep{SDSS1, SDSS2} of the Sloan Digital Sky Survey (SDSS), the number
of known satellites increased significantly thanks to new discoveries
made with the survey. Table~\ref{tab:measured_satellites} includes the
new satellite galaxies found with SDSS \citep{SatData3,SatData4,
  SatData5, SatData6, SatData7, SatData8, SatData9, SatData10,
  SatData11, SatData12, SatData16, SatData17, SatData18, SatData19},
using the recently published structural parameters given in
\citet{SatData13}. To estimate the half light radius of the Large
Magellanic Cloud, the formula $r_h = 1.68 \, r_e$ \citep{SatData13} was
adopted.

\begin{figure*}
	\centering
  \begin{minipage}[t]{0.47\textwidth}
    \centering
    \includegraphics[width=\textwidth]{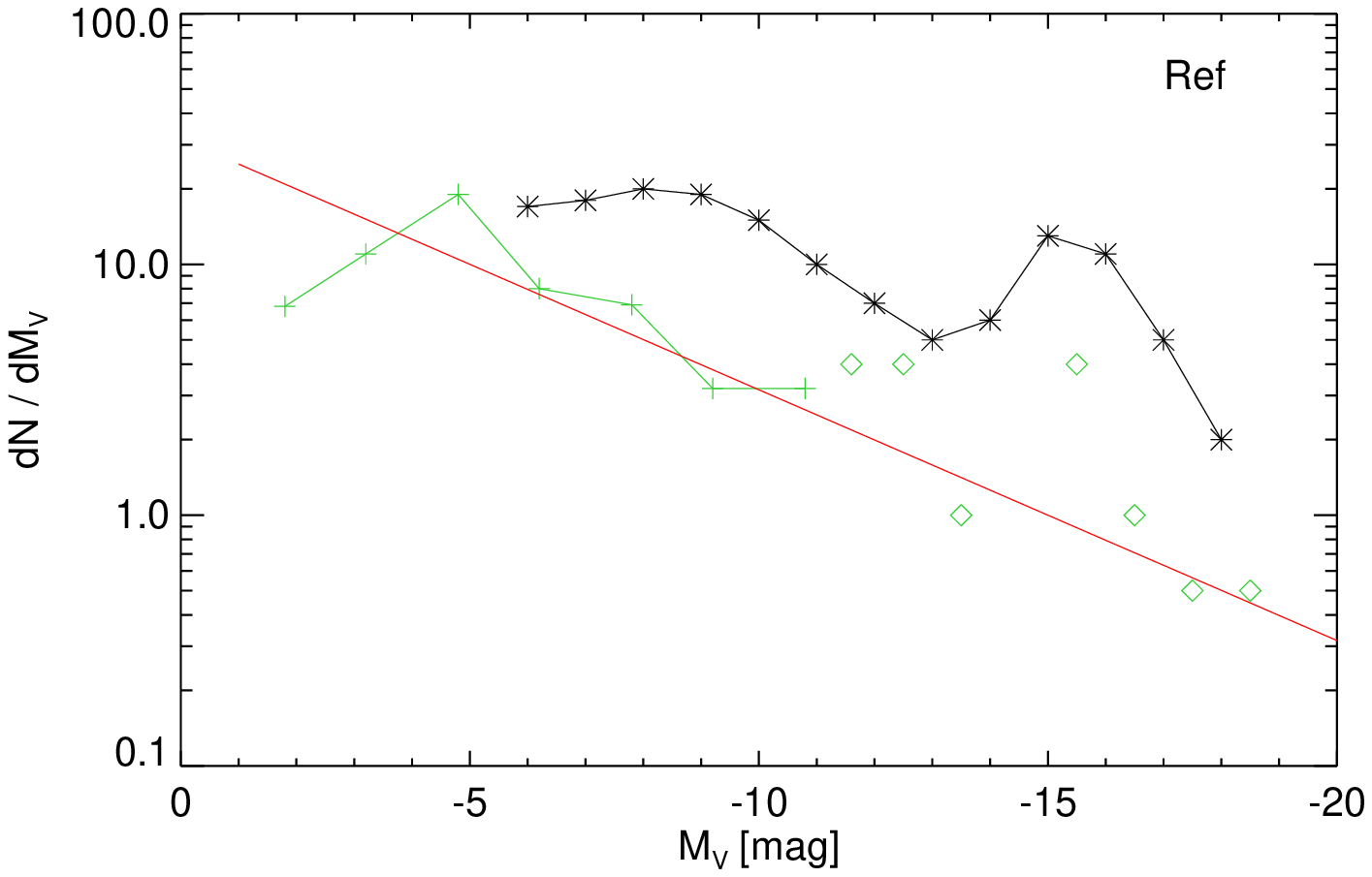} 
  \end{minipage}
  \begin{minipage}[t]{0.47\textwidth}
    \centering
    \includegraphics[width=\textwidth]{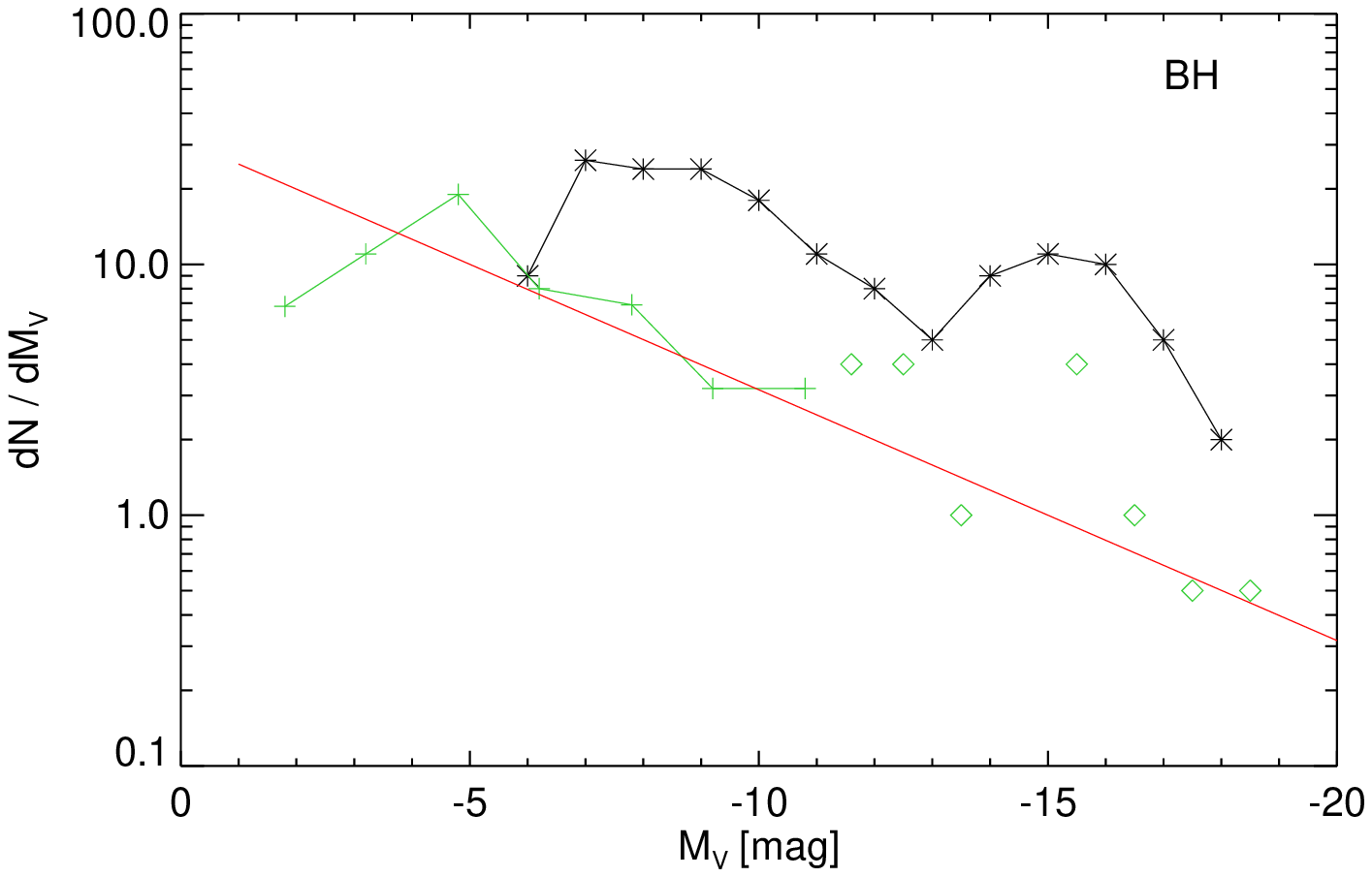} 
  \end{minipage}
  \hfill
  \begin{minipage}[t]{0.47\textwidth}
    \centering
    \includegraphics[width=\textwidth]{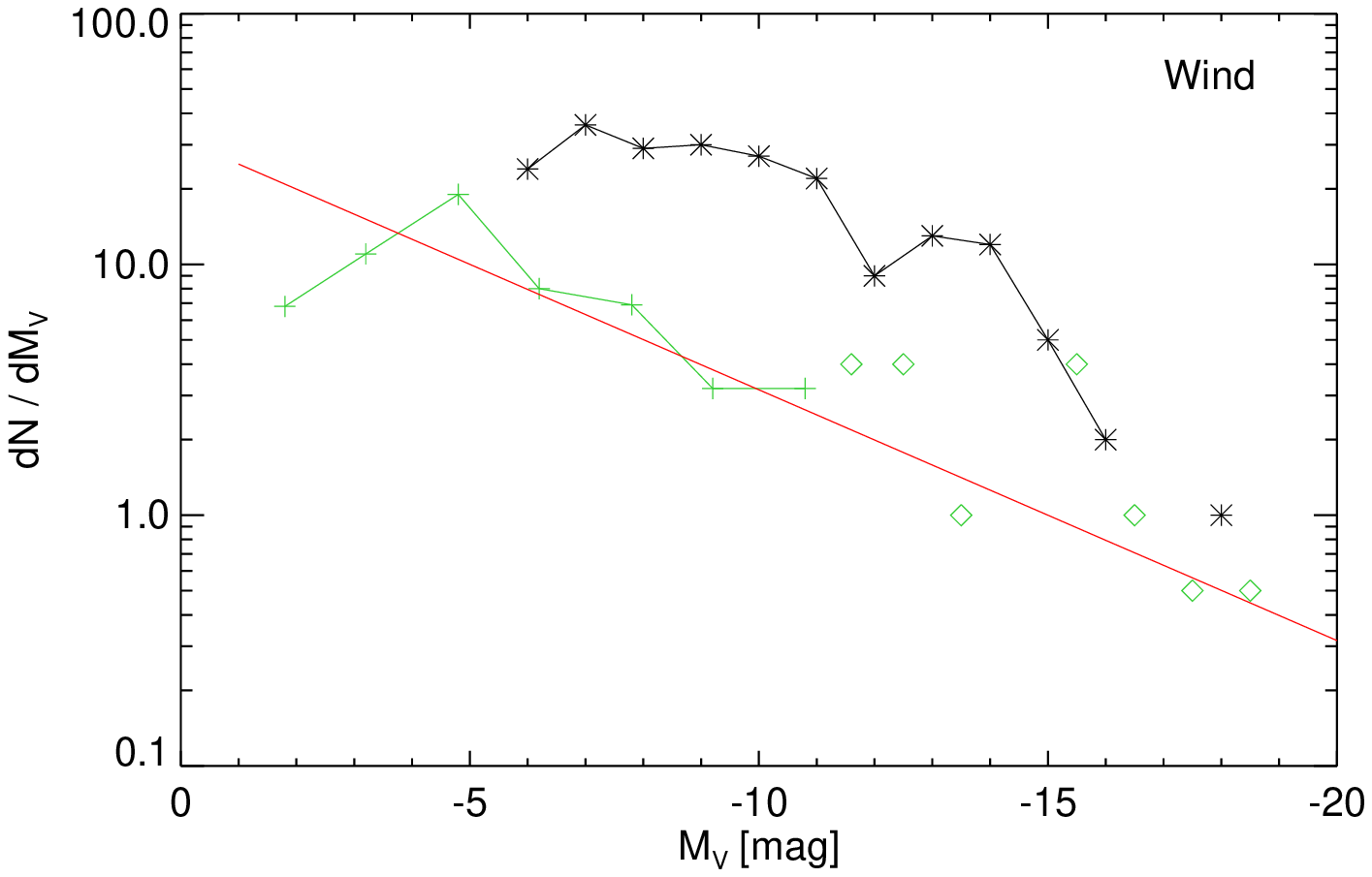} 
  \end{minipage}
  \begin{minipage}[t]{0.47\textwidth}
    \centering
    \includegraphics[width=\textwidth]{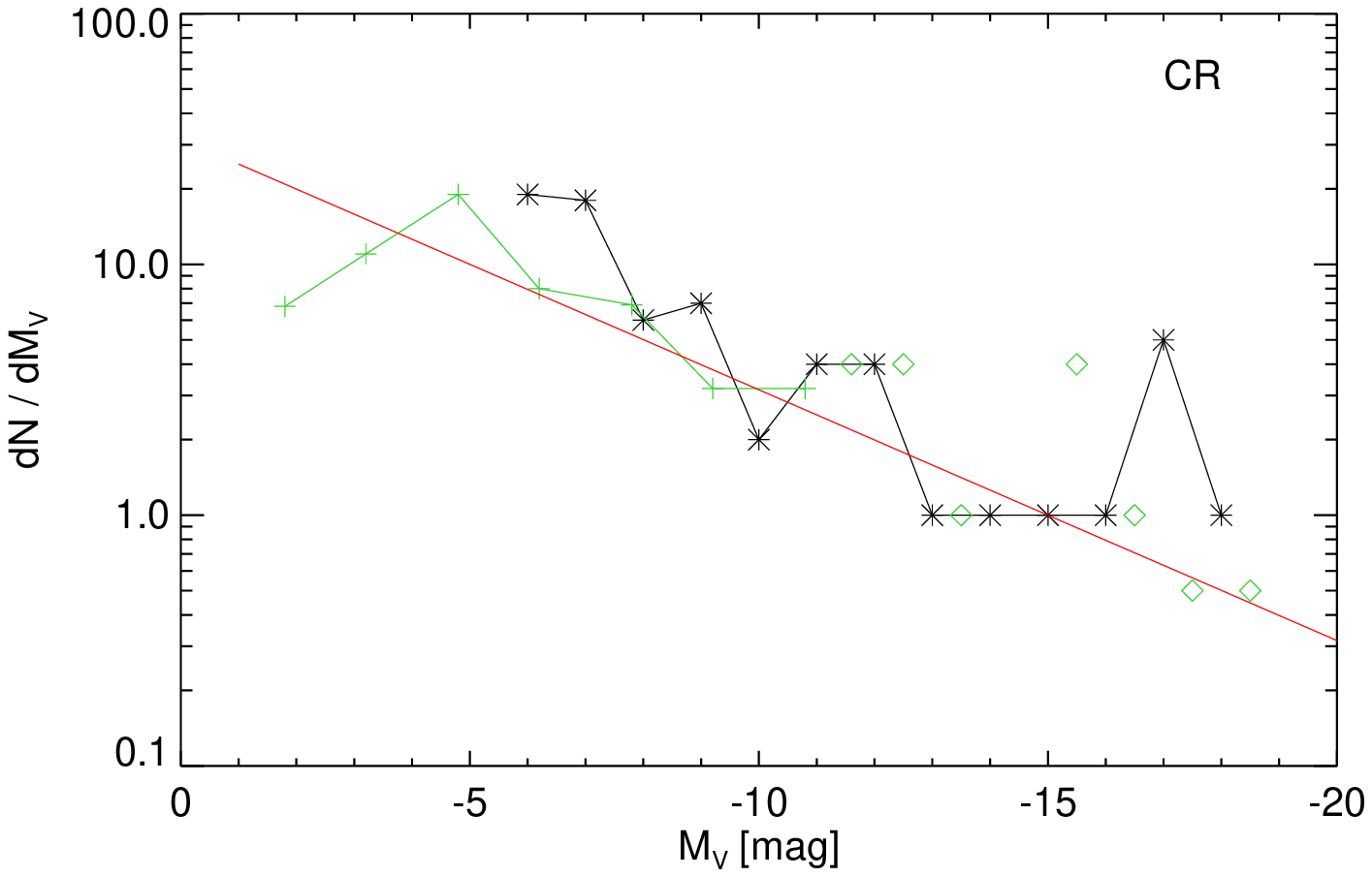} 
  \end{minipage}
  \hfill
	\caption{The differential luminosity function of the simulated
          satellites (black curve) compared to the observational data.
          The latter are represented by the red line, which is the
          fitting function given by \citet{Koposov08}. The green line
          shows the scaled luminosity function they obtained for the
          SDSS satellites while the open diamonds give an extended
          version using also the luminous satellites of the Milky Way
          and Andromeda. The different panels show
           our results for the \textsc{Ref}, \textsc{BH},
          \textsc{Wind} and \textsc{CR} simulations, respectively.}%
	\label{fig:Satellites_Luminosity_Function}%
\end{figure*}

Up to now, there are $35$ known Milky Way satellites. However, this
sample is not complete, as the recently found satellites are all limited
to the area of the sky covered by SDSS, which corresponds to a fraction
of $0.194$ of the full sky.  Effectively, only this region of the whole
sky has been scanned for faint satellites
\citep[see][]{HundredSatellites}. Taking into account the detection
limits and the sky coverage of the SDSS survey, \citet{SatKinematics}
estimate the number of satellite galaxies with a surface brightness
above $\approx 28 \; \mathrm{mag/arcsec^2}$ \citep{SatData13} expected
over the whole sky to be $57$. However, more recent works favor a limit
of $30\; \mathrm{mag/arcsec^2}$ \citep{Bullock2009}, which we adopt
throughout the rest of this paper. Using this threshold, we denote
simulated satellite galaxies as 'observable' if their surface brightness
exceeds $30\; \mathrm{mag/arcsec^2}$.

\section{Abundance of luminous satellites}\label{sec:abundance}

Arguably the most fundamental property of the subhalo population is the
abundance of satellites as a function of luminosity.  In
Figure~\ref{fig:Satellites_Luminosity_Function}, we show the differential
luminosity function of the simulated satellite galaxies for our four
primary simulations, and compare to the observations for the Milky
Way. The latter are expressed in terms of the fitting formula given by
\citet{Koposov08}:

\begin{equation}
	\frac{\text{d}N}{\text{d}M_V} = 10 \times 10 ^{0.1(M_V + 5)}.
\end{equation}

The upper left
panel of Figure~\ref{fig:Satellites_Luminosity_Function} shows our
result for the reference simulation. There is a sizable offset between
the simulated and the observed luminosity functions, shown in black and
green, respectively. Satellites with large stellar masses are even more
strongly overproduced than low luminosity ones. This shows that
photoheating and supernova feedback as included in the reference model
are insufficient to match the observed satellite abundance.

The other panels of the figure show the results we obtained for our
alternative physics simulations.  As one might expect, including
supermassive black holes has no substantial influence on the population
of satellites, yielding essentially the same result as for our reference
simulation. This shows that environmental heating effects from quasar
feedback, in particular the possible quenching of star formation in
nearby small halos, play no important role in the history of the Aq-C
halo. Interestingly, galactic outflows with our standard wind
prescription are also unable to significantly improve the agreement with
the observations. While the most luminous satellites are moderately
suppressed in stellar mass, the satellites tend to pile up on the faint
side of the luminosity function, yielding to a slight steepening effect
of the luminosity function, quite different from what is needed to match
the data. Finally, the simulation including cosmic rays yields a
substantial modification of the results. Here we actually obtain very
good agreement with the luminosity function inferred from the
observations, because compared with the reference model the luminosity
of the satellites is efficiently suppressed by the CR feedback.

\begin{figure}
	\centering%
	\includegraphics[width = \linewidth]{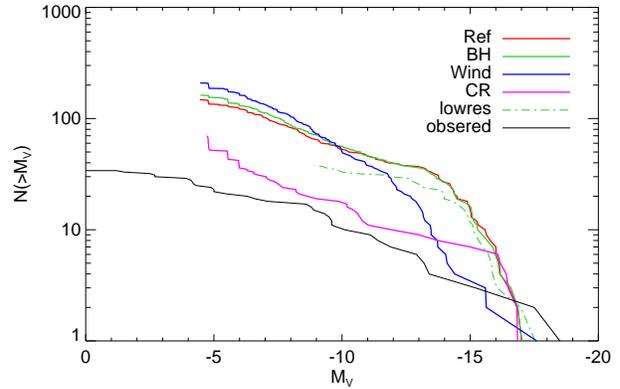}%
	\caption{Cumulative satellite luminosity function for our
          different simulation models, compared to the observed
          luminosity function.  Note that the observed luminosity
          function in this plot includes only known satellites without
          any incompleteness correction and should thus be seen as a
          lower limit, especially below $\approx -7\;M_V$ where
          satellites can only be detected within the $19.4 \%$ sky
          coverage of SDSS.}
	\label{fig:Satellites_Luminosity_Function_cum}%
\end{figure}

Another view on the satellite abundance is given in
Figure~\ref{fig:Satellites_Luminosity_Function_cum}, where we show the
cumulative abundance of the satellite population as a function of
luminosity, comparing all four simulation results to the observations.
Here, the observational data is based only on direct observations,
meaning that the black line should be taken as a lower limit for
magnitudes lower than $\approx -7\;M_V$, because of the incomplete sky
coverage of the SDSS. The figure confirms the conclusions we reached
from the different results of
Fig.~\ref{fig:Satellites_Luminosity_Function}. The `\textsc{Wind}'
simulation is efficient in reducing the luminosity of large satellites
of the size of the LMC/SMC, but does not manage to suppress the
abundance of low luminosity satellites. In contrast, while the CR
simulation does not reduce the luminosity of the brightest satellites
much, it is very efficient in suppressing star formation in low-mass
subhalos, ultimately producing a much reduced amplitude of the
luminosity at the faint end. As a result, the CR simulation stays quite
close to the observational data up to the completeness limit of the SDSS.

Figure~\ref{fig:Satellites_Luminosity_Function_cum} also includes the
result of the `\textsc{LowRes}' simulation, shown as a green dot-dashed
line. We can see that the population of satellite galaxies is
independent of resolution to good accuracy up to a magnitude of $\approx
-12\;M_V$. Taking into account that the resolution limit of our high
resolution runs is shifted by $\approx 4$ magnitudes, we expect that our
simulated satellite abundance should be numerically converged for
satellites brighter than $\approx -8\;M_V$.
 
\begin{figure*}
	\centering
  \begin{minipage}[t]{0.47\textwidth}
    \centering
    \includegraphics[width=\textwidth]{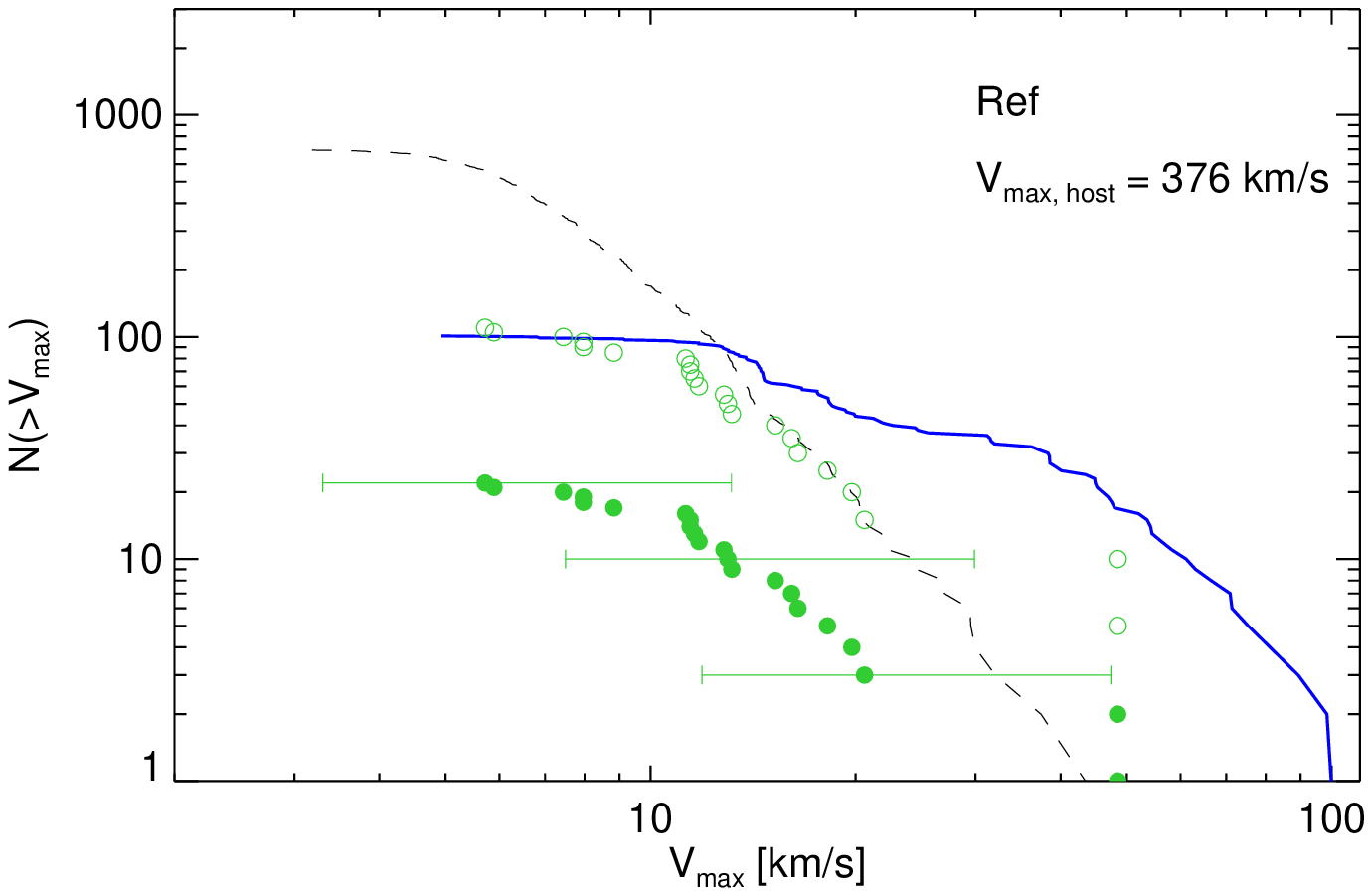} 
  \end{minipage}
  \begin{minipage}[t]{0.47\textwidth}
    \centering
    \includegraphics[width=\textwidth]{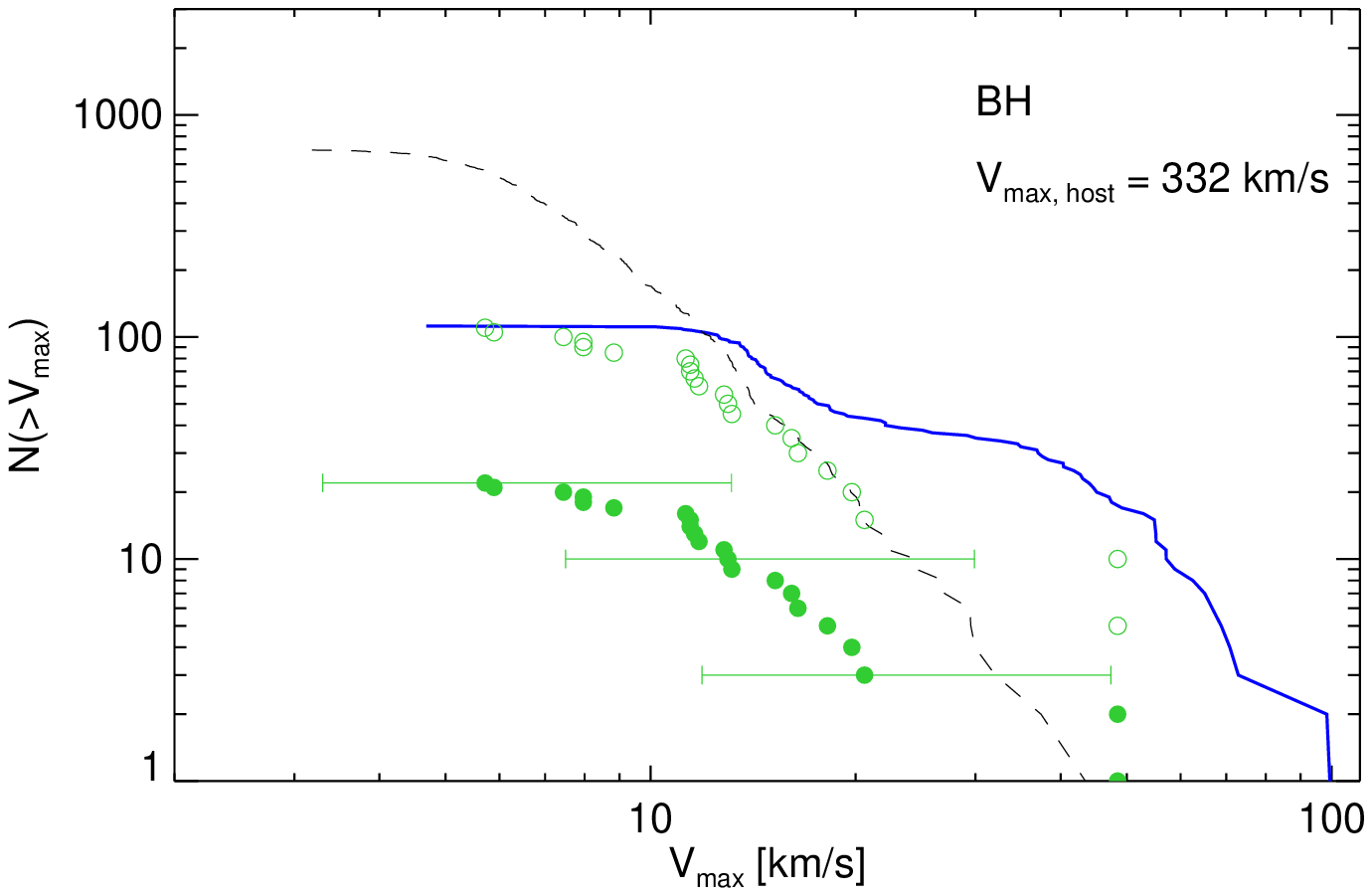} 
  \end{minipage}
  \hfill
  \begin{minipage}[t]{0.47\textwidth}
    \centering
    \includegraphics[width=\textwidth]{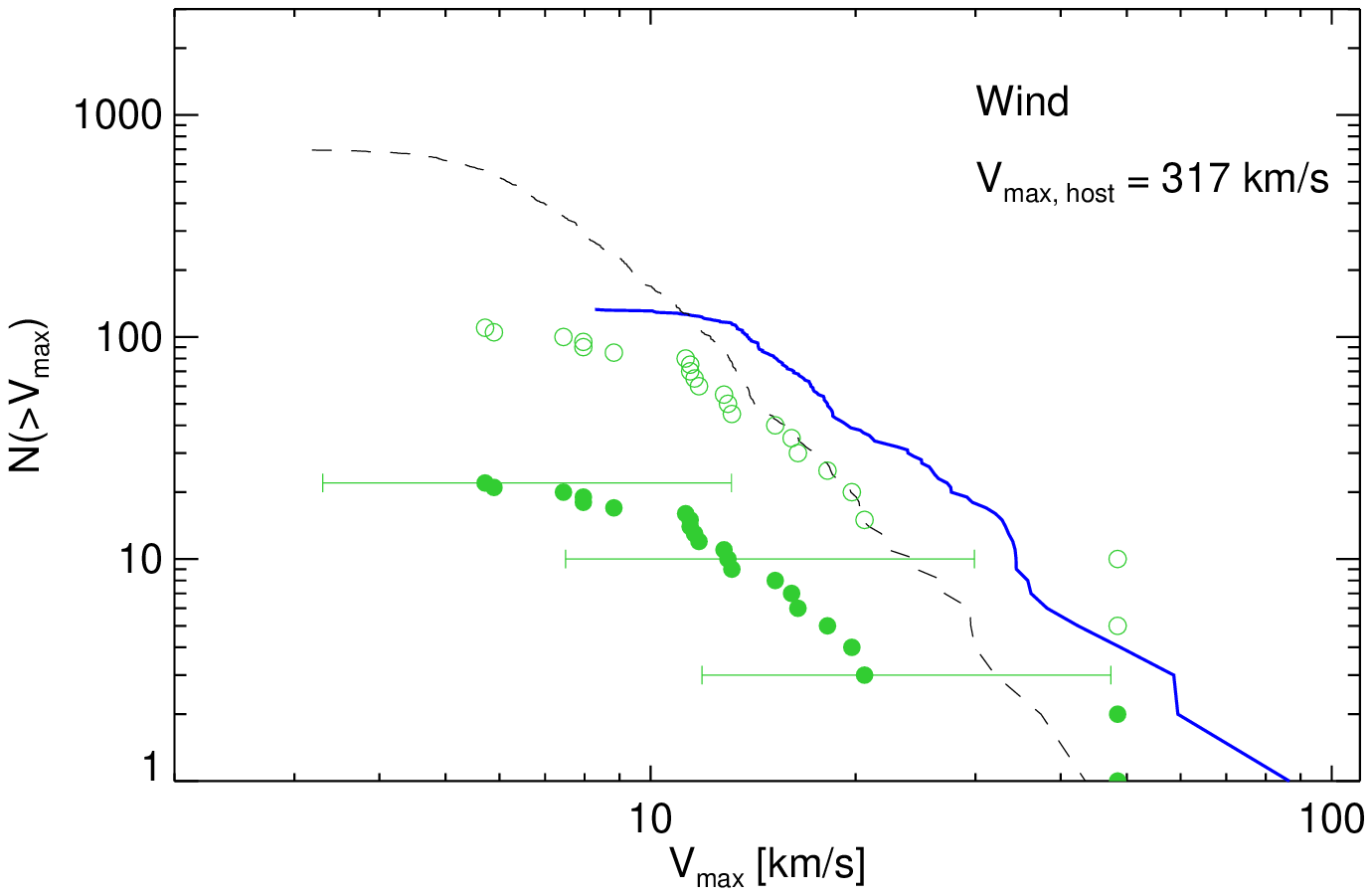} 
  \end{minipage}
  \begin{minipage}[t]{0.47\textwidth}
    \centering
    \includegraphics[width=\textwidth]{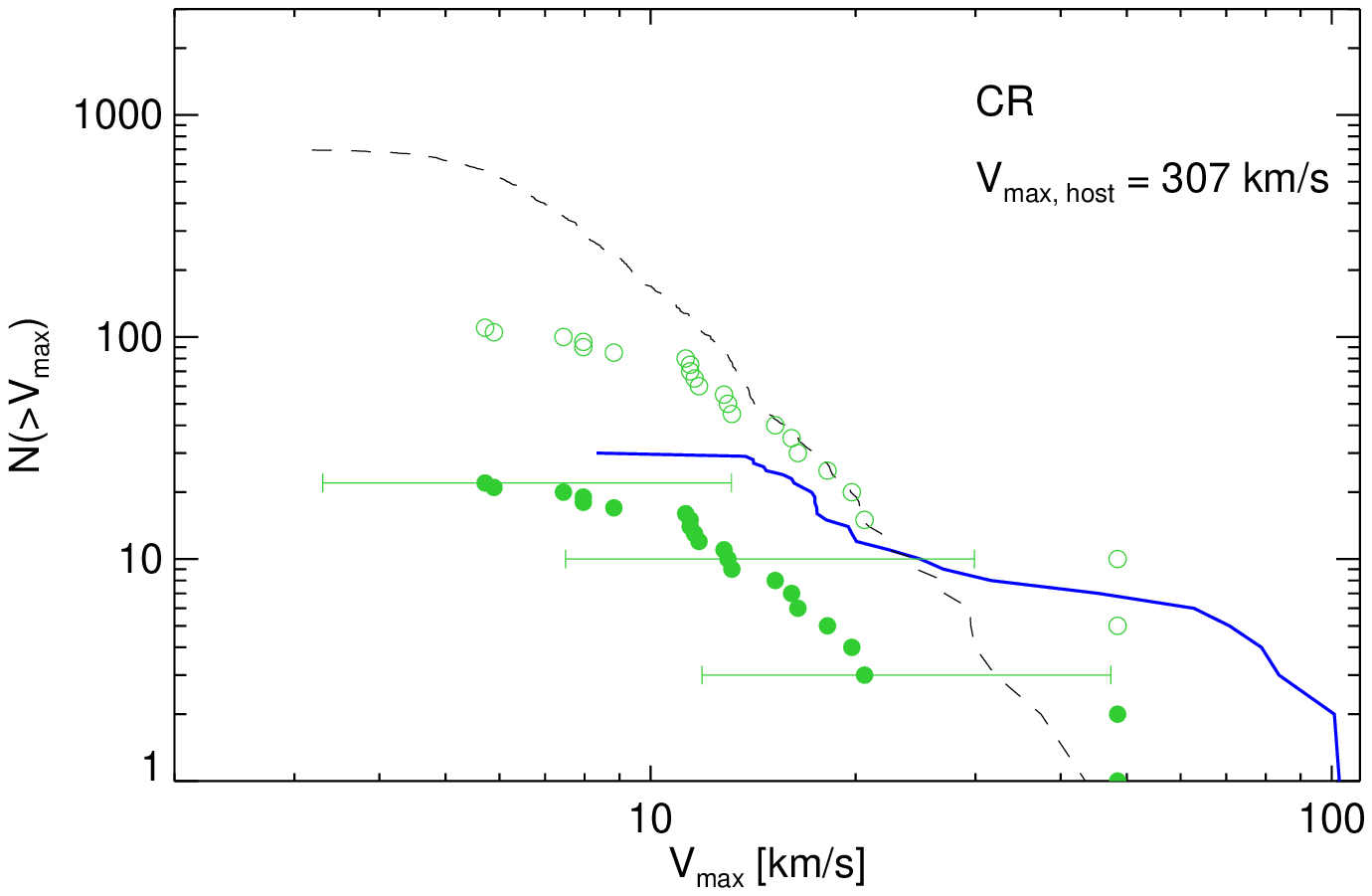} 
  \end{minipage}
  \hfill
	\caption{The cumulative number of luminous satellites (with
          stellar mass larger than $10^5\,h^{-1}{\rm M}_\odot$) as
          function of their maximum circular velocity. We compare
          results for our different physics simulations with data for
          the Milky Way satellites, taken from \citet{Via_Lactea_III,
            SatKinematics, SatData2, Martin07} and plotted as green
          filled circles together with the assumption that the circular
          velocity equals $\sqrt{3}$ times the central velocity
          dispersion \citep{Primack09}.  The error bars indicate a
          plausible range of circular velocities between $\sigma$ and
          $4\times\sigma$. Open circles show the same data as the solid
          circles, but scaled by a factor of $5$ to roughly account for
          SDSS sky coverage and incompleteness.  The blue line shows the
          cumulative mass function of the observable satellites in each
          hydrodynamical simulation, while for comparison the black
          dashed line gives the mass function of all substructures in
          the corresponding dark matter only simulation. The variations
          in the maximum circular velocity obtained for the host halos
          in the different simulations are due to the different sizes of
          the stellar bulges grown in the different runs.}%
	\label{fig:Satellites_CumNbr}%
\end{figure*}

Figure~\ref{fig:Satellites_CumNbr} shows yet another way to compare the
counts of simulated satellites with the observations. Here we use the
maximum circular velocity of satellites, $v_{\mathrm{max}}$, on the
abscissa, because circular velocities are a good proxy for the
(original) mass of the subhalos, but can be much more reliably measured
than the mass itself. We note that such velocity functions have already
been used in the first discussions of the missing satellite problem, and
are still frequently used to compare the number of observed satellites
with the substructure abundance in collisionless N-body simulations
\citep[e.g.][]{Via_Lactea_III}. The filled green circles in
Fig.~\ref{fig:Satellites_CumNbr} show the raw observational data, while
the open circles are a scaled version that accounts for the SDSS sky
coverage and incompleteness. The real cumulative velocity function might
thus be expected to lie close to the filled symbols at high circular
velocities and to approach the open circles at low circular
velocities. The dashed black line shows the cumulative velocity function
of all satellite galaxies produced in a dark matter only simulation,
using the same initial conditions as for the high resolution
hydrodynamical simulations. Finally, the blue line in each panel shows
the cumulative mass function of all satellites containing at least
$1\times 10^4 \,h^{-1}{\rm M_{\odot}}$ of stellar mass (one star
particle) in the corresponding hydrodynamic simulation.

The differences between the observations and the simulation results for
the different physics models appear large at first sight. This however
confirms and is consistent with our earlier findings. In particular, the
reference simulation and the simulation with black hole feedback
overpredict the satellite counts for all velocities, while the wind
simulation at least manages to give a reasonable abundance of the
brightest satellite systems. Again, we find the cosmic ray simulation to
produce the best match to the data. Whereas there may still be a
moderate overproduction of bright systems, the extrapolated faint end
abundance is matched quite well, and, in particular, the shape of the
predicted luminosity function is in quite good agreement with the
observations.

There is another interesting aspect of
Figure~\ref{fig:Satellites_CumNbr} that concerns the comparison with the
dark matter only results. It is a generally assumed that satellite
galaxies are dark matter dominated. However, comparing the black dashed
line, which shows the mass function of satellites in the corresponding
dark matter only simulation starting from the same initial conditions,
with the result of the individual hydrodynamic simulations, we note some
sizable differences. The high mass satellites show clear evidence that
gas cooling has led to a higher concentration of their mass profiles,
thereby increasing their circular velocities. Despite the relatively low
stellar mass content in these bright satellites, they hence show some
structural changes due to baryonic effects. We note however that
invoking yet stronger supernovae feedback may reduce these effects if
the cooling rate is more effectively reduced.

\subsection{Kinematic results}

\begin{figure}
	\centering%
	\includegraphics[width =\linewidth]{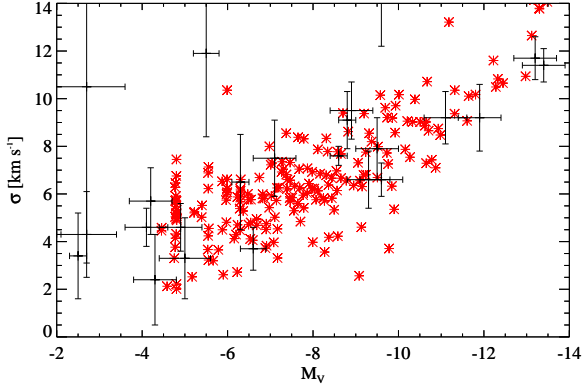}%
	\caption{Total V-band luminosity against velocity dispersion.
          We show the observational data points of \citet{SatKinematics}
          combined with velocity dispersions taken from
          \citet{UniversalMassProfile} as black crosses, and compare to
          the simulated satellites plotted as red symbols.  The two
          samples are in good agreement, except perhaps for slightly
          different slopes.
\label{fig:Satellites_Vmag_sigma}}%
\end{figure}

We close this section with an anlysis of some of the structural
properties of the simulated satellite population. As noted earlier, the
total mass of a satellite galaxy is difficult to measure, so other
tracers are usually used as a proxy for mass.  An observationally
readily accessible measure of this type is the central velocity
dispersion, which is very commonly used \citep[e.g.][]{SatKinematics}.
In Figure~\ref{fig:Satellites_Vmag_sigma}, we compare the relation
between central velocity dispersion and luminosity for our simulation
satellites with data from \citet{SatKinematics}, updated with the latest
values for the known satellites from \citet{UniversalMassProfile}.

There seems to be quite good agreement between the observations and the
dark matter velocity dispersions of the \textsc{Ref} simulation. As the
sample of measured satellites is quite small and has rather large error
bars, the weak trend of rising velocity dispersion with rising
luminosity is not very well determined, but the simulation apparently
follows the same trend. We note that the alignment of the simulated
satellites on the left hand side of the plot is due to resolution issues
from discreteness effects. In fact, the different `stripes' are
separated by just one star particle. The stellar mass in each stripe is
therefore equal, even though some scatter in luminosity is still present
because the luminosity was calculated with the \citet{Bruzual} model,
taking into account the age and metallicity of the star particles,
effectively giving each stellar particle its own mass-to-light ratio.

\section{History of satellite galaxies}\label{sec:history}

\begin{figure*}
	\centering
  \begin{minipage}[t]{0.32\textwidth}
    \centering
    \includegraphics[width=\textwidth]{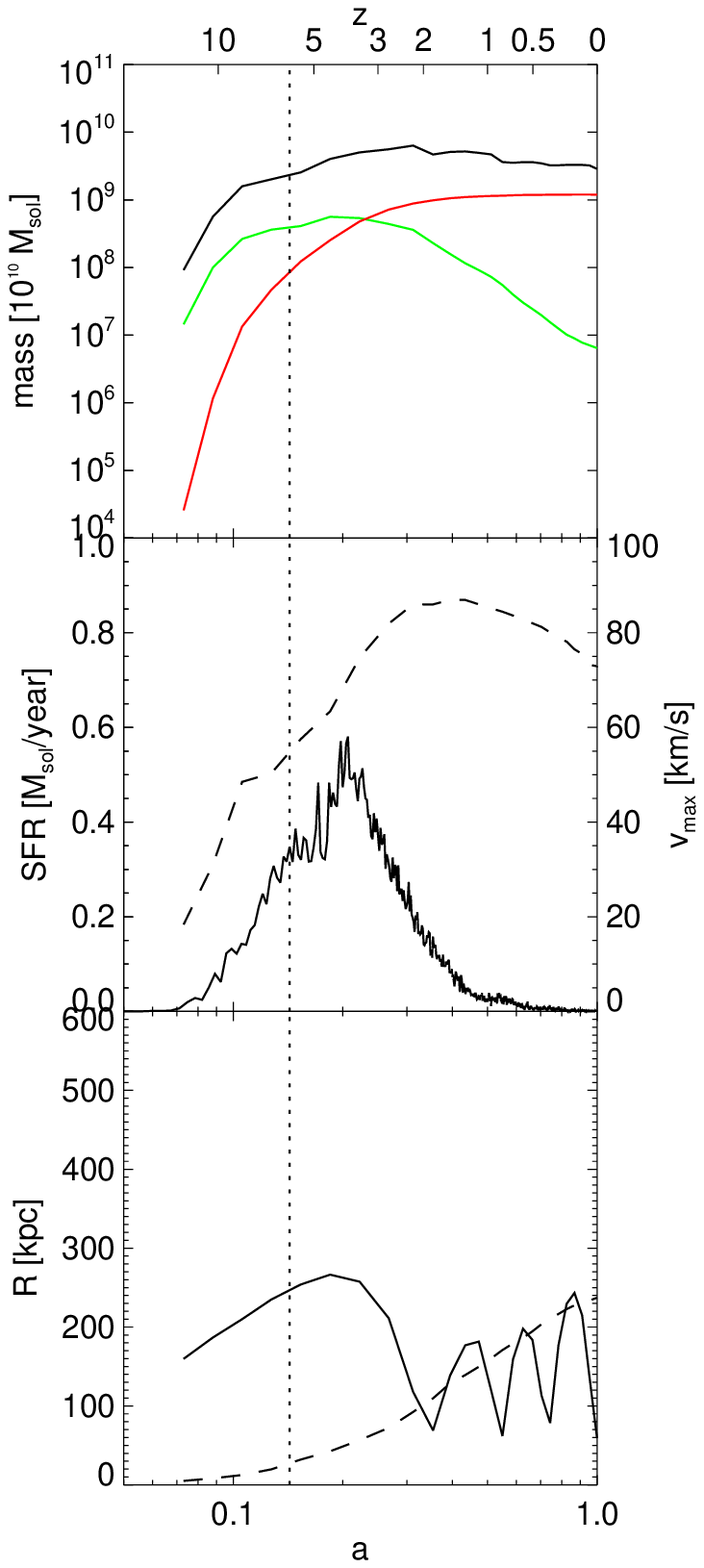} 
  \end{minipage}
  \begin{minipage}[t]{0.32\textwidth}
    \centering
    \includegraphics[width=\textwidth]{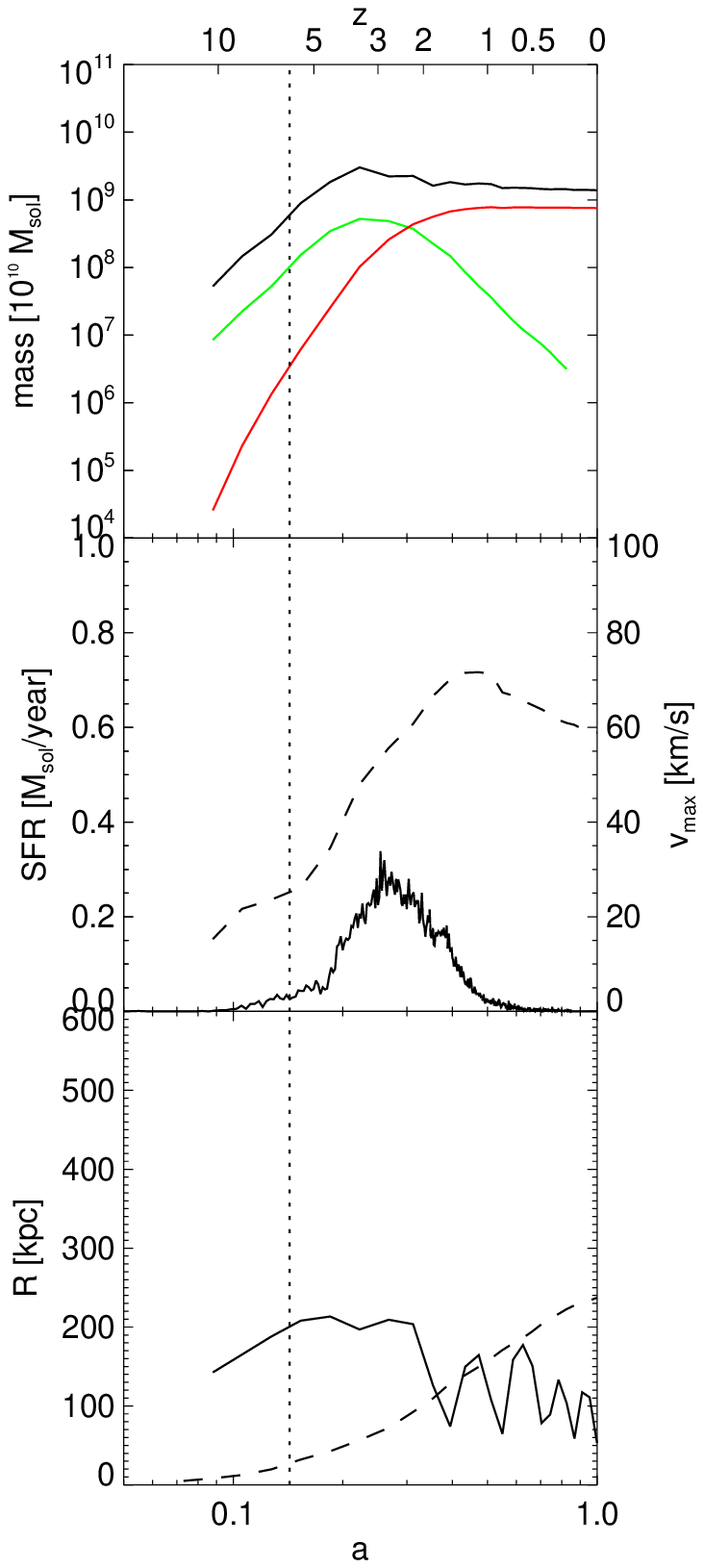} 
  \end{minipage}
    \begin{minipage}[t]{0.32\textwidth}
    \centering
    \includegraphics[width=\textwidth]{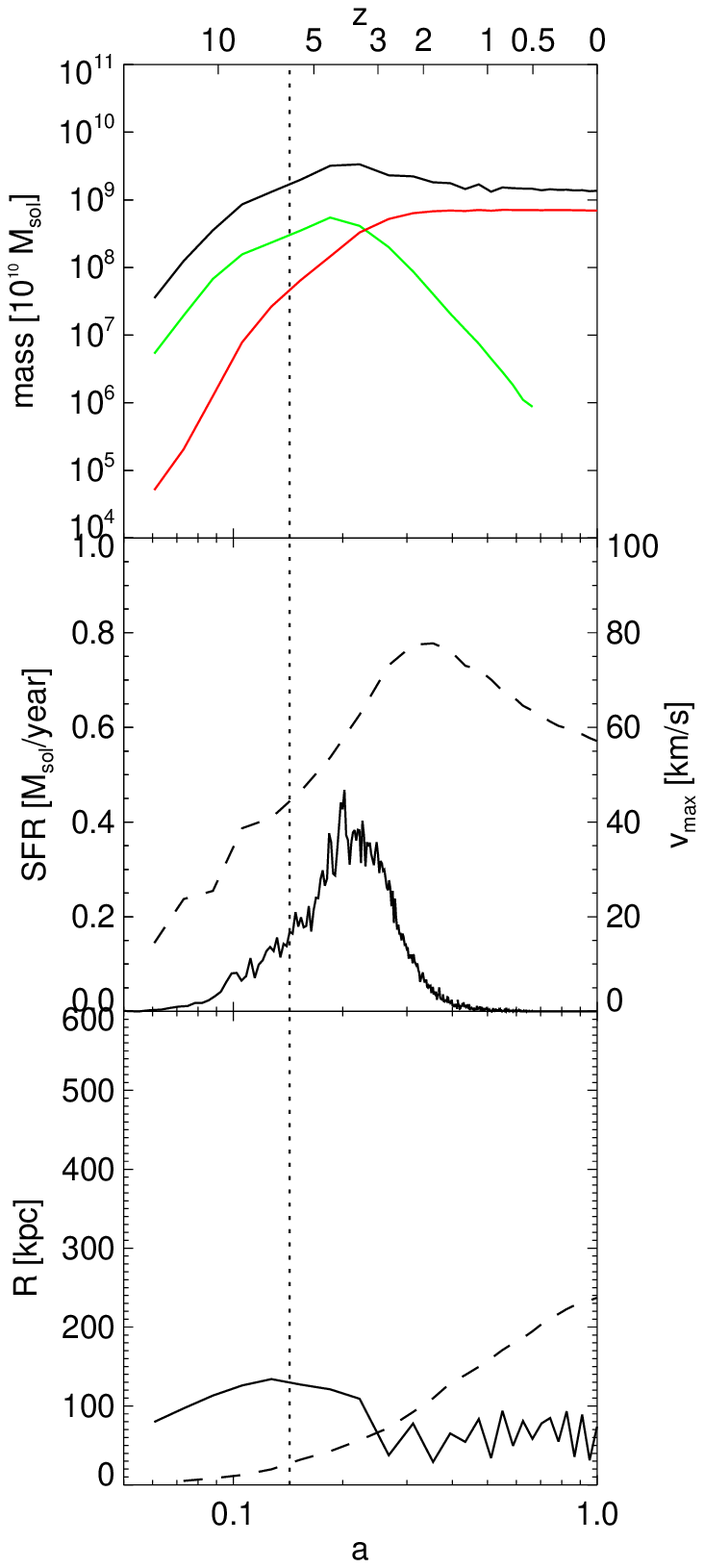} 
  \end{minipage}
	\caption{Detailed time evolution of three representative
          examples of high-mass satellites, from the time of their first
          appearance as individual galaxies until $z=0$. In each case,
          the top panel shows the evolution of the dark matter, gas and
          stellar mass components as black, green and red lines,
          respectively. The middle panel shows the star formation rate
          as solid line, and the maximum circular velocity as a dashed
          line (with the corresponding scale on the right $y$-axis). The
          bottom panel gives the radial distance of the satellite to the
          host galaxy at each timestep (solid line), and compares this
          to the virial radius ($R_{200}$) shown with a dashed line. The
          dotted vertical lines in all panels mark the $z=6$ epoch of
          reionization in our simulations.
	\label{fig:Satellites_high_mass_bin}}
\end{figure*}

\begin{figure*}
	\centering
  \begin{minipage}[t]{0.32\textwidth}
    \centering
    \includegraphics[width=\textwidth]{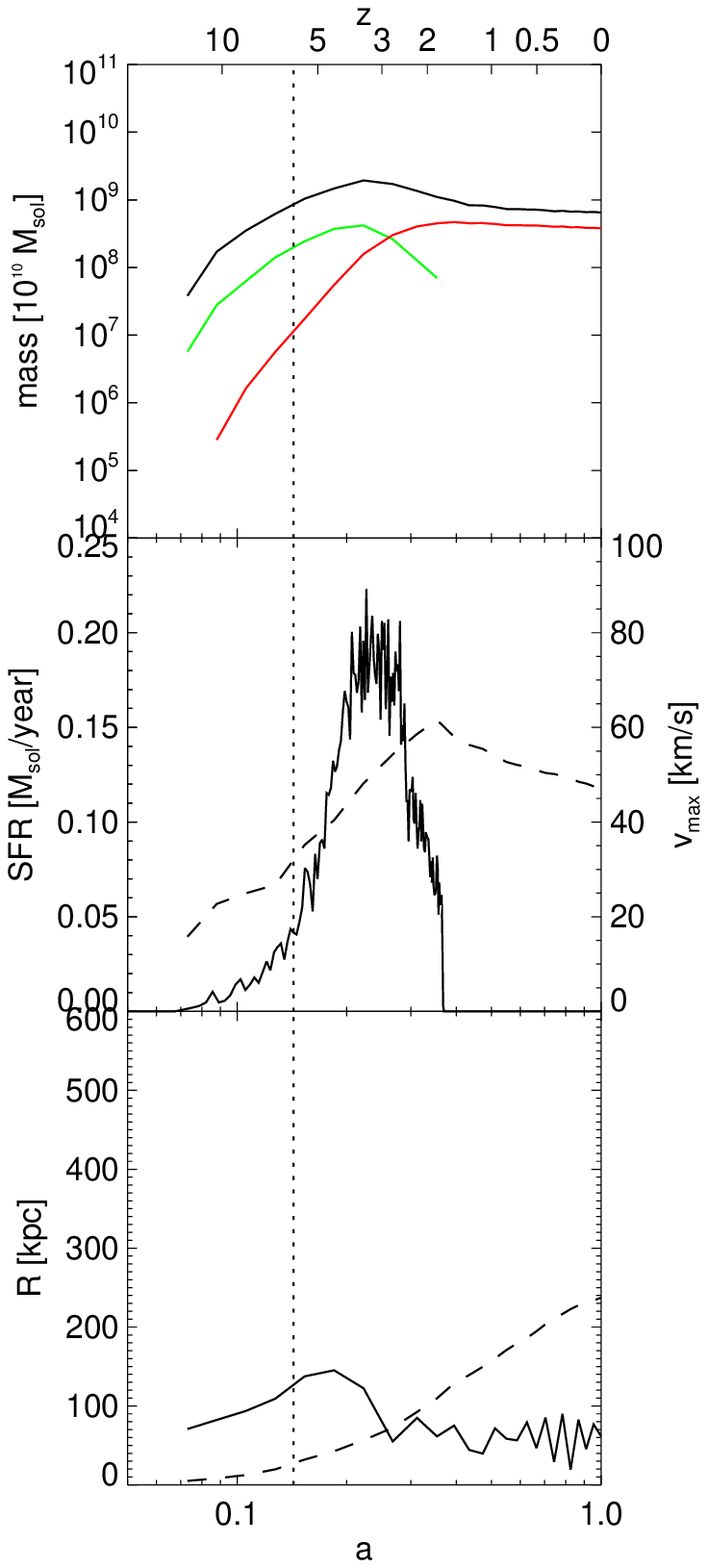} 
  \end{minipage}
  \begin{minipage}[t]{0.32\textwidth}
    \centering
    \includegraphics[width=\textwidth]{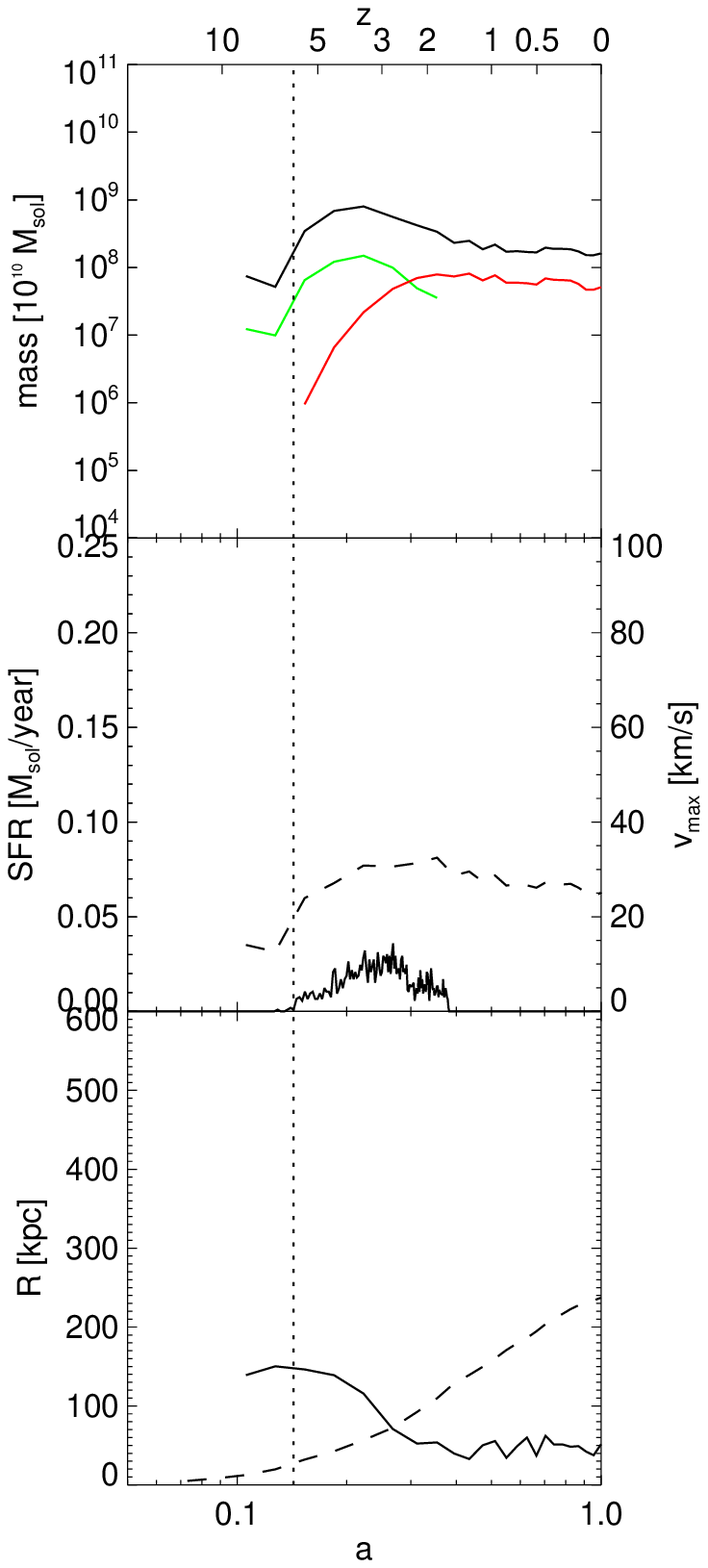} 
  \end{minipage}
    \begin{minipage}[t]{0.32\textwidth}
    \centering
    \includegraphics[width=\textwidth]{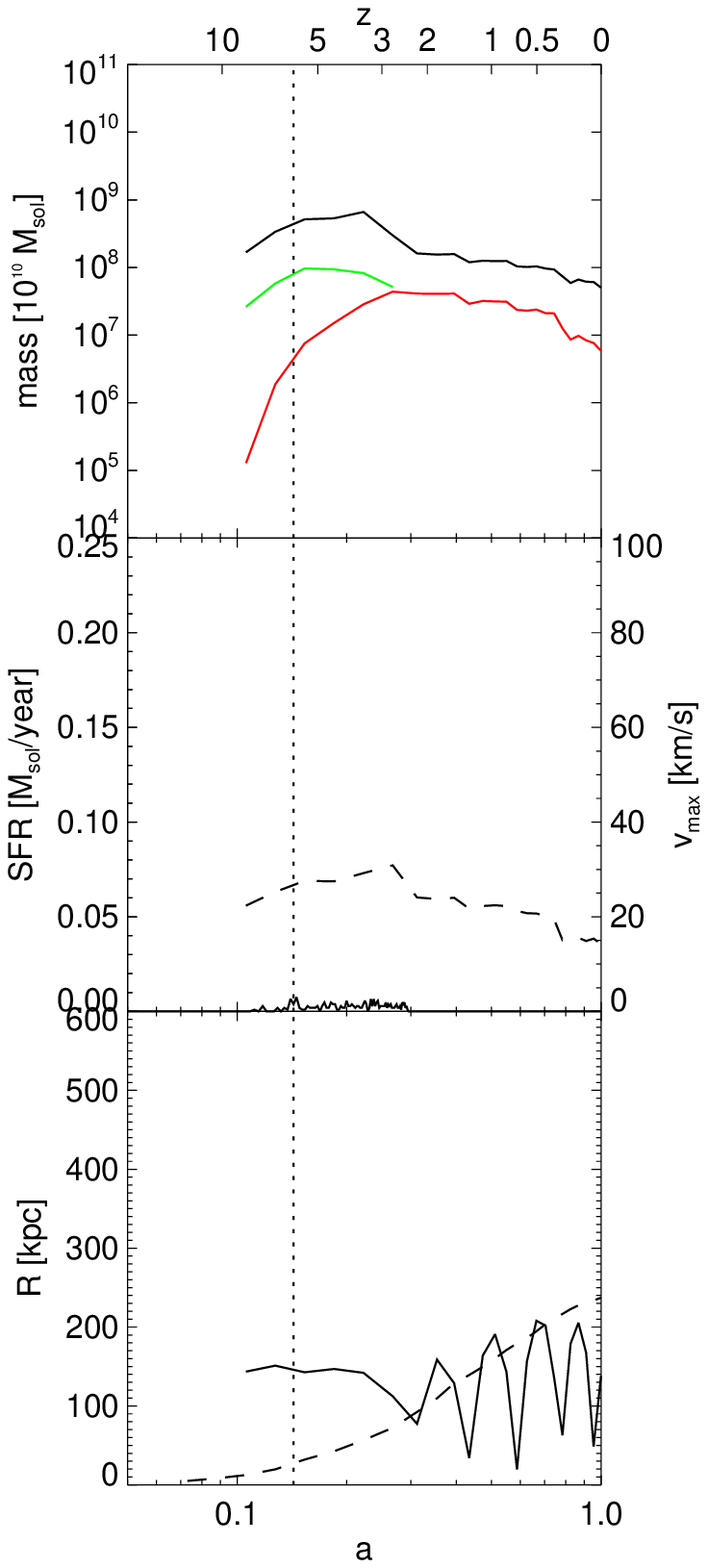} 
  \end{minipage}
	\caption{The same as Fig.~\ref{fig:Satellites_high_mass_bin},
          but for three intermediate mass
          satellites. \label{fig:Satellites_medium_mass_bin}}
\end{figure*}

\begin{figure*}
	\centering
  \begin{minipage}[t]{0.32\textwidth}
    \centering
    \includegraphics[width=\textwidth]{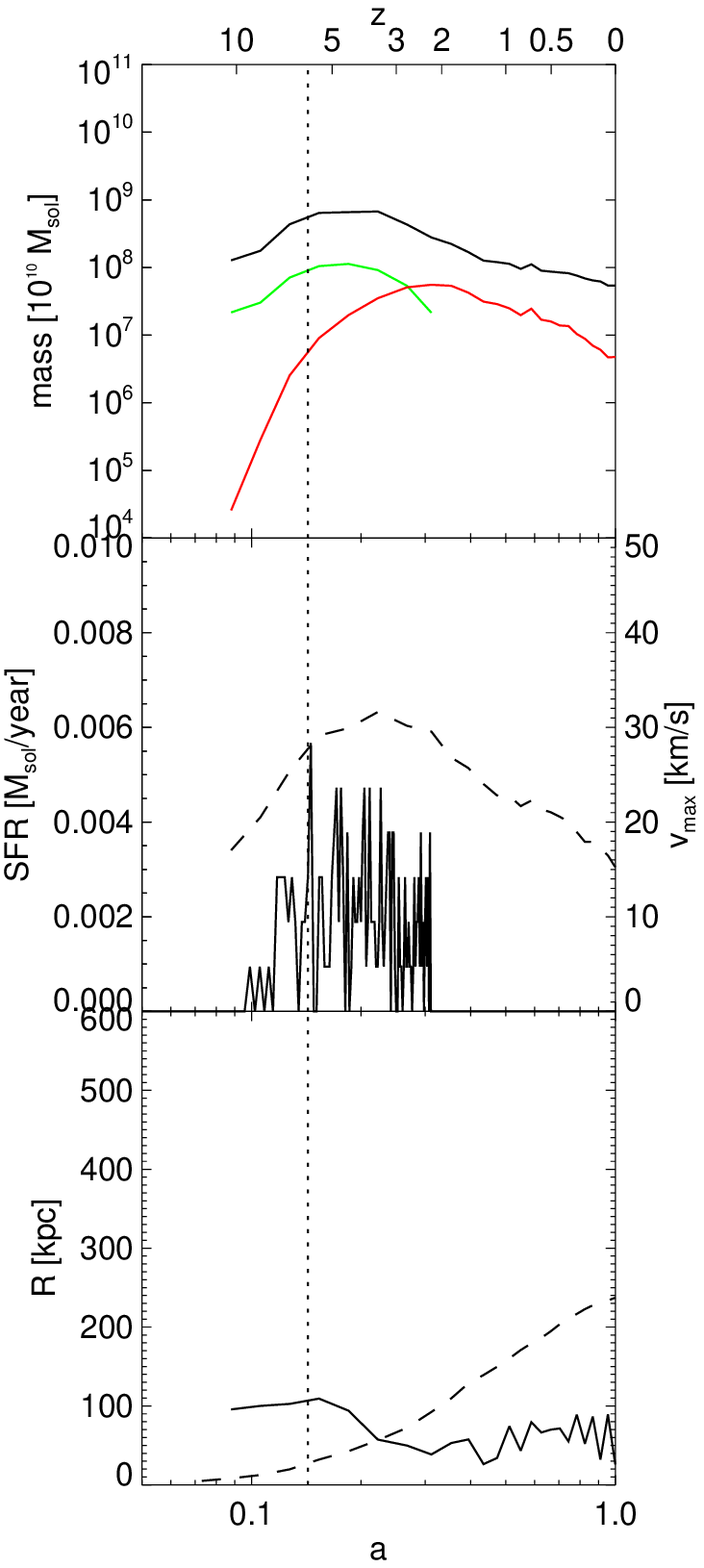} 
  \end{minipage}
  \begin{minipage}[t]{0.32\textwidth}
    \centering
    \includegraphics[width=\textwidth]{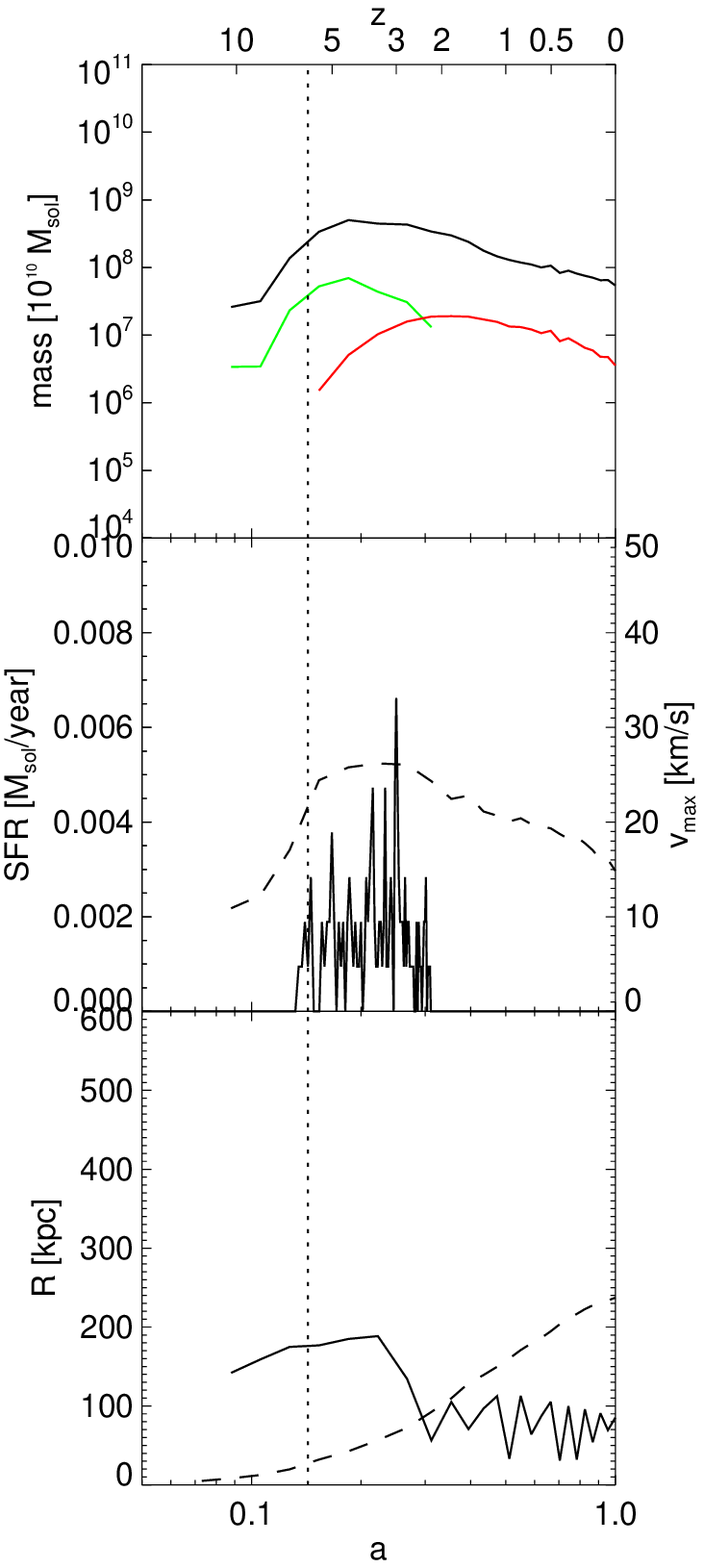} 
  \end{minipage}
    \begin{minipage}[t]{0.32\textwidth}
    \centering
    \includegraphics[width=\textwidth]{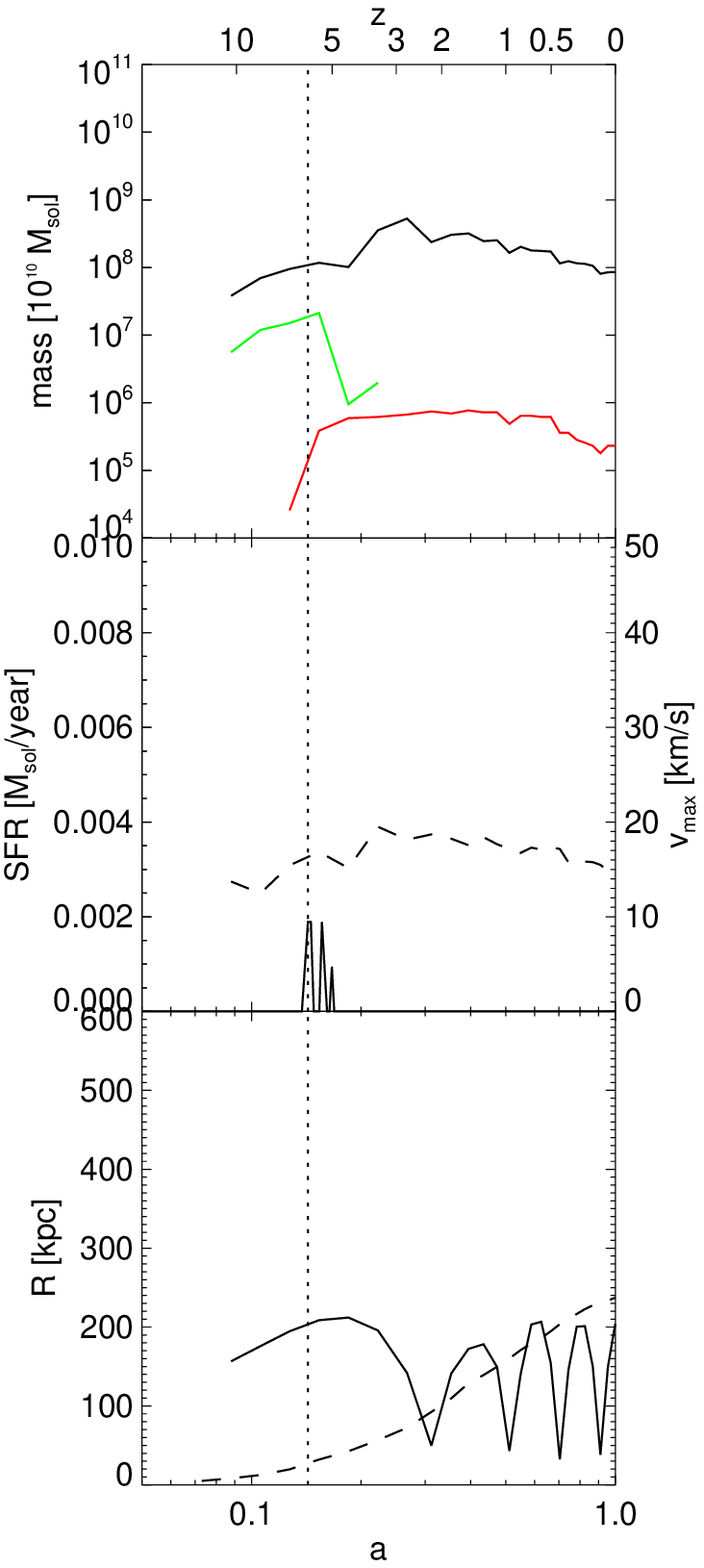} 
  \end{minipage}
	\caption{The same as Fig.~\ref{fig:Satellites_high_mass_bin} 
  but  for three low mass satellites.}%
	\label{fig:Satellites_low_mass_bin}%
\end{figure*}

In this section we track the evolution of individual satellites, with
the aim to study their formation paths for a range of individual
accretion, mass loss and star formation histories. To this end we select
nine representative satellite galaxies, split into groups of three that
are taken from three different mass ranges. In~Figure
\ref{fig:Satellites_high_mass_bin}, we show our `high mass sample',
consisting of three satellite galaxies with a final stellar mass higher
than $5 \times 10^8\;\mathrm{M_\odot}$. Satellites with intermediate
final stellar mass between $5 \times 10^8\;\mathrm{M_\odot}$ and $5
\times 10^6\;\mathrm{M_\odot}$ are shown in
Figure~\ref{fig:Satellites_medium_mass_bin}, while
Figure~\ref{fig:Satellites_low_mass_bin} gives three low-mass examples
of satellites with a final stellar mass less than $5 \times
10^6\;\mathrm{M_\odot}$.

The different panels in the three Figures
\ref{fig:Satellites_high_mass_bin} to \ref{fig:Satellites_low_mass_bin}
are organized in the same way, and show in each case the history of one
individual satellite (with a final stellar mass as labeled in the
figure). For each satellite, the top panel gives the redshift evolution
of the dark matter, gas and stellar components as black, green and red
solid lines, respectively. The middle panel shows both the star
formation rate (solid line) and the maximum circular velocity (dashed
line), as a function of time. Finally, the bottom panel gives the
evolution of the radial distance of the satellite to the host galaxy
(solid line) and compares this to the virial radius of the host
($R_{200}$, dashed line). Finally, the dotted vertical line running
through all panels highlights the epoch $z=6$, which is the time when
the UV background reionizes the universe in our simulations.

\begin{figure}
    \includegraphics[width=\linewidth]{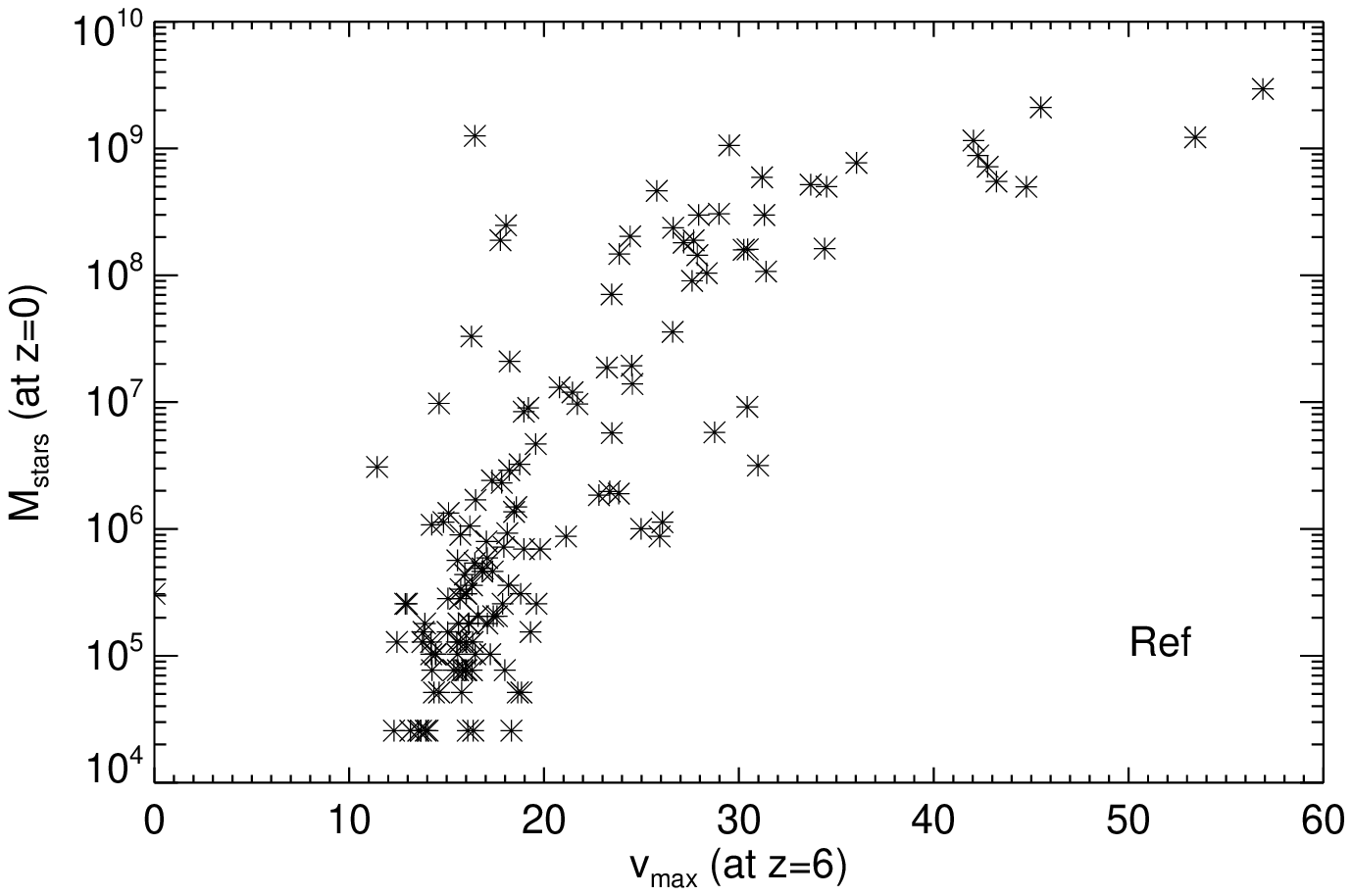} 
    \includegraphics[width=\linewidth]{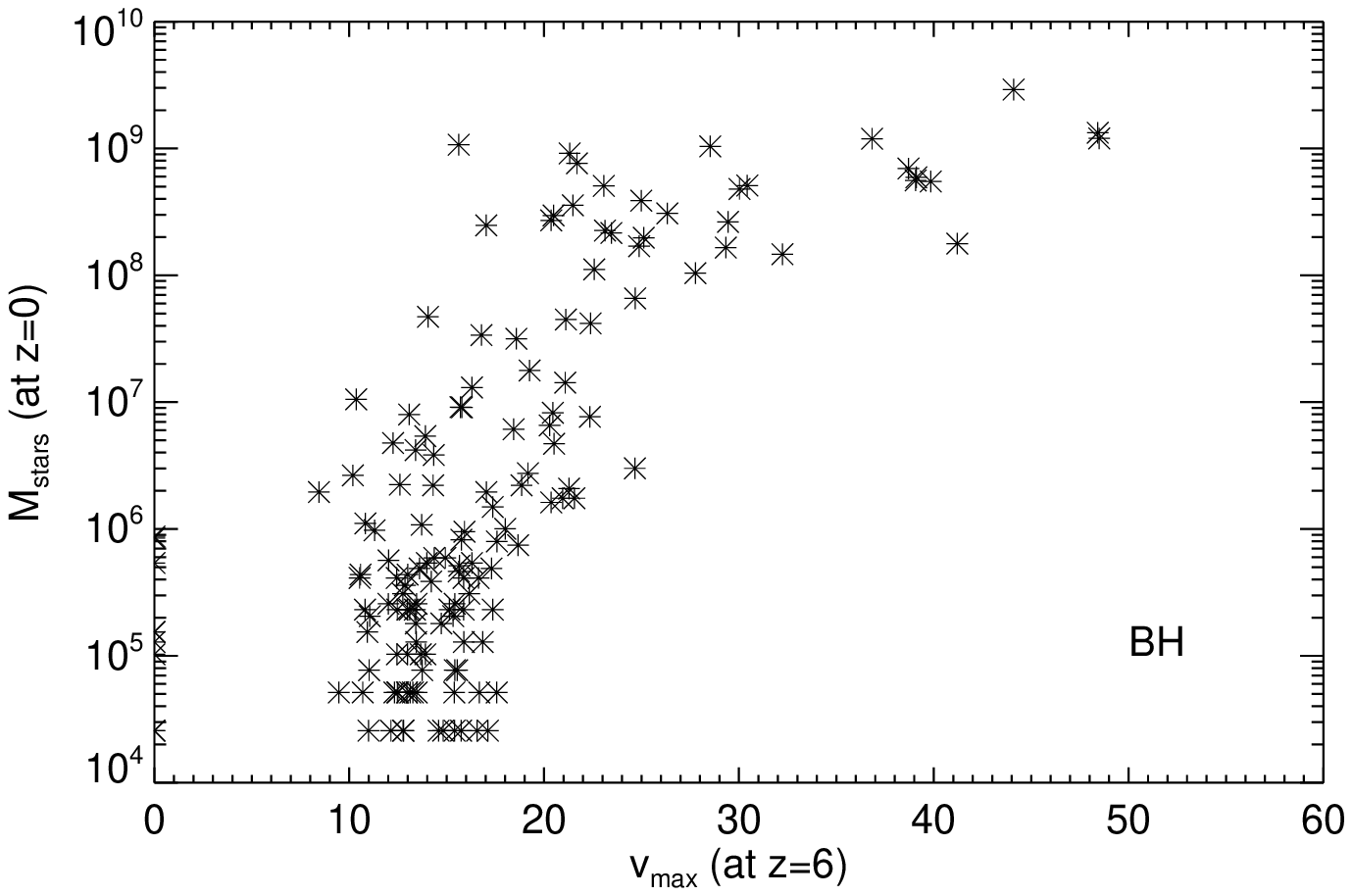} 
    \includegraphics[width=\linewidth]{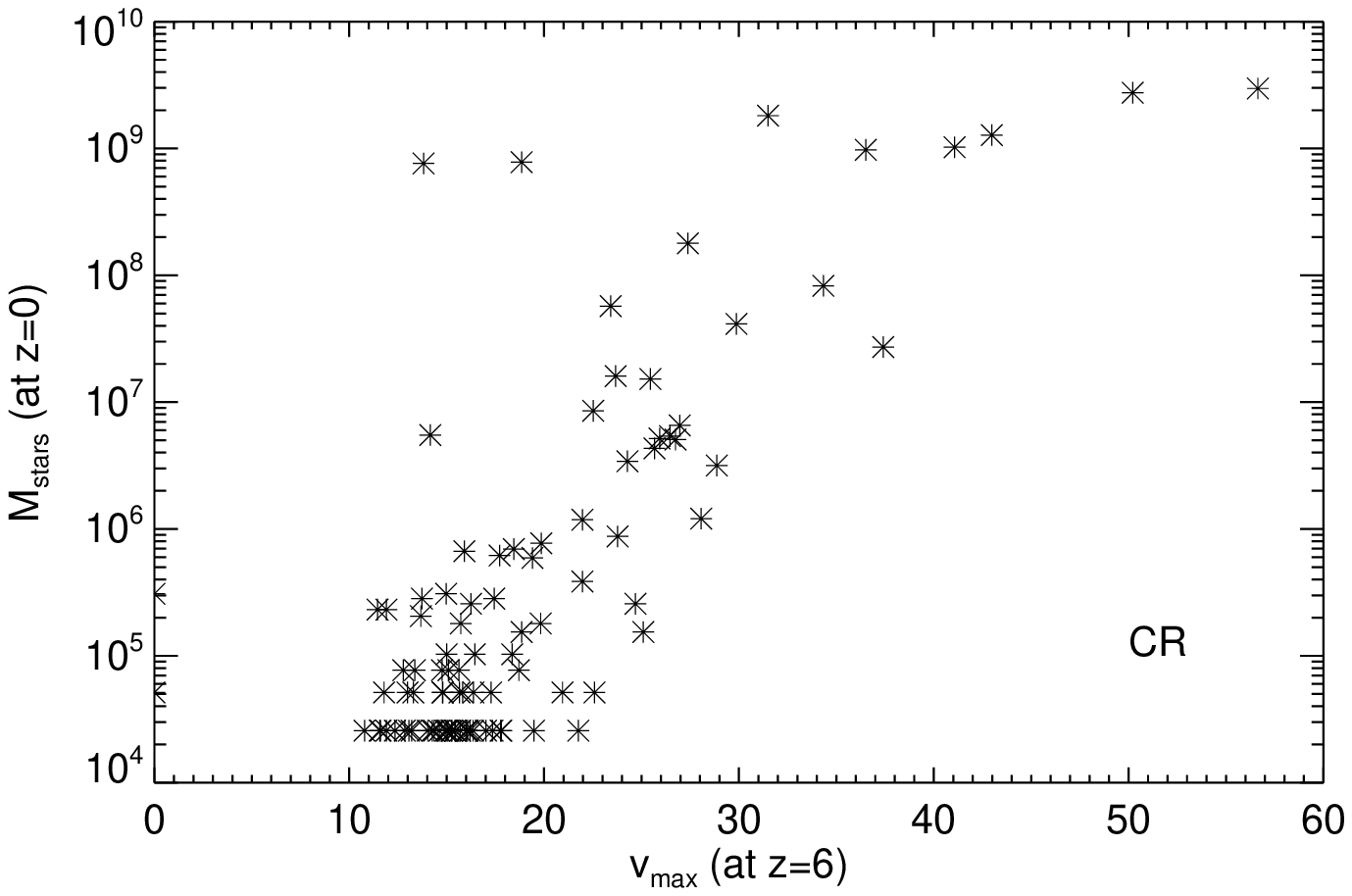} 
	\caption{Current stellar mass of satellites versus their maximum
          circular velocity at $z = 6$. We do not find a clear threshold
          in $v_{\mathrm{max}}$ at $ z = 6$ that could be used to decide
          whether or not a satellite galaxy is able to form stars later
          on. Instead, we find that the final stellar mass shows large
          scatter over a considerable range of circular velocities at
          the epoch of reionization.
\label{fig:vmax_at_reio}}%
\end{figure}

\begin{figure}
    \includegraphics[width=\linewidth]{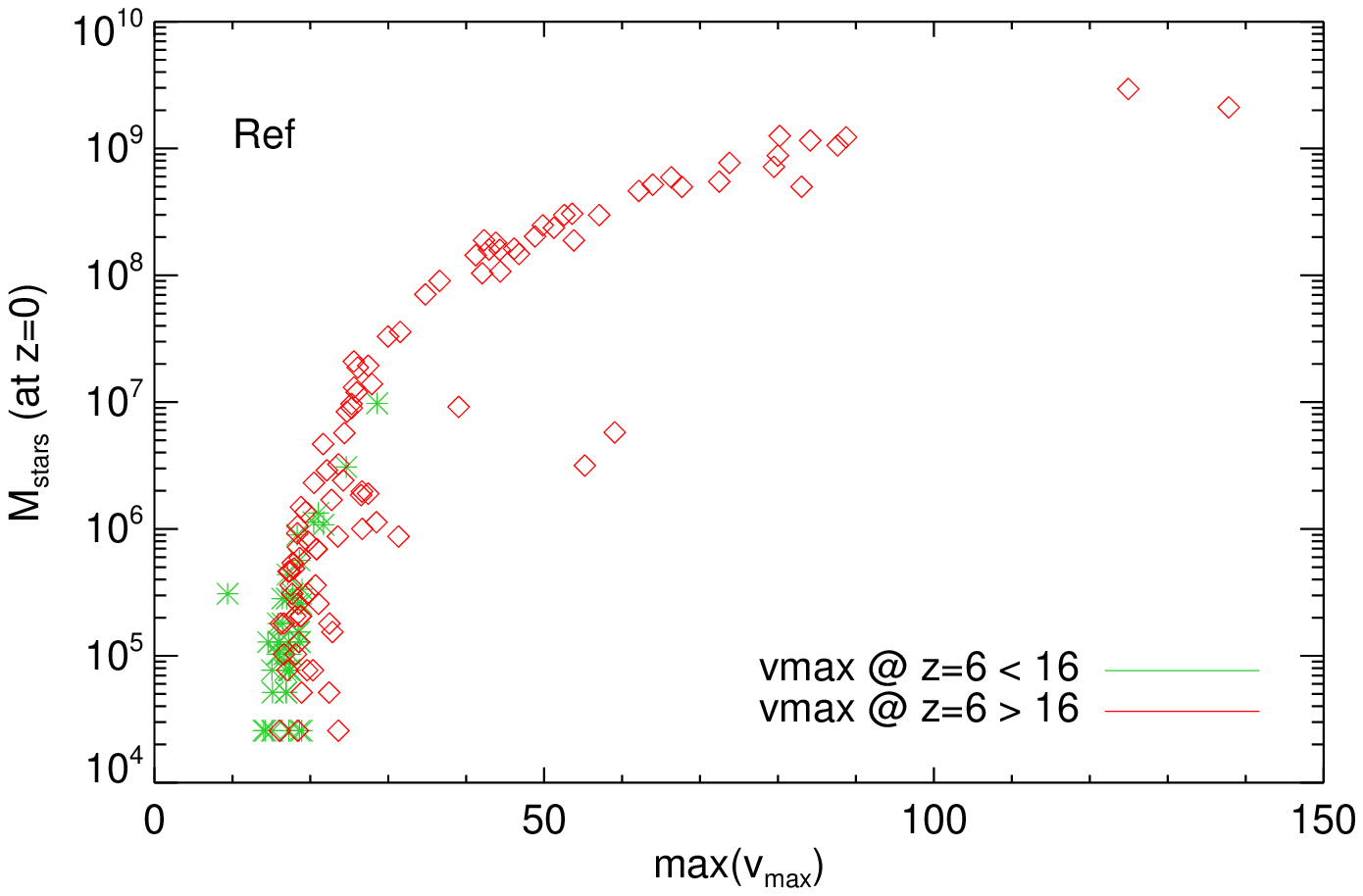} 
    \includegraphics[width=\linewidth]{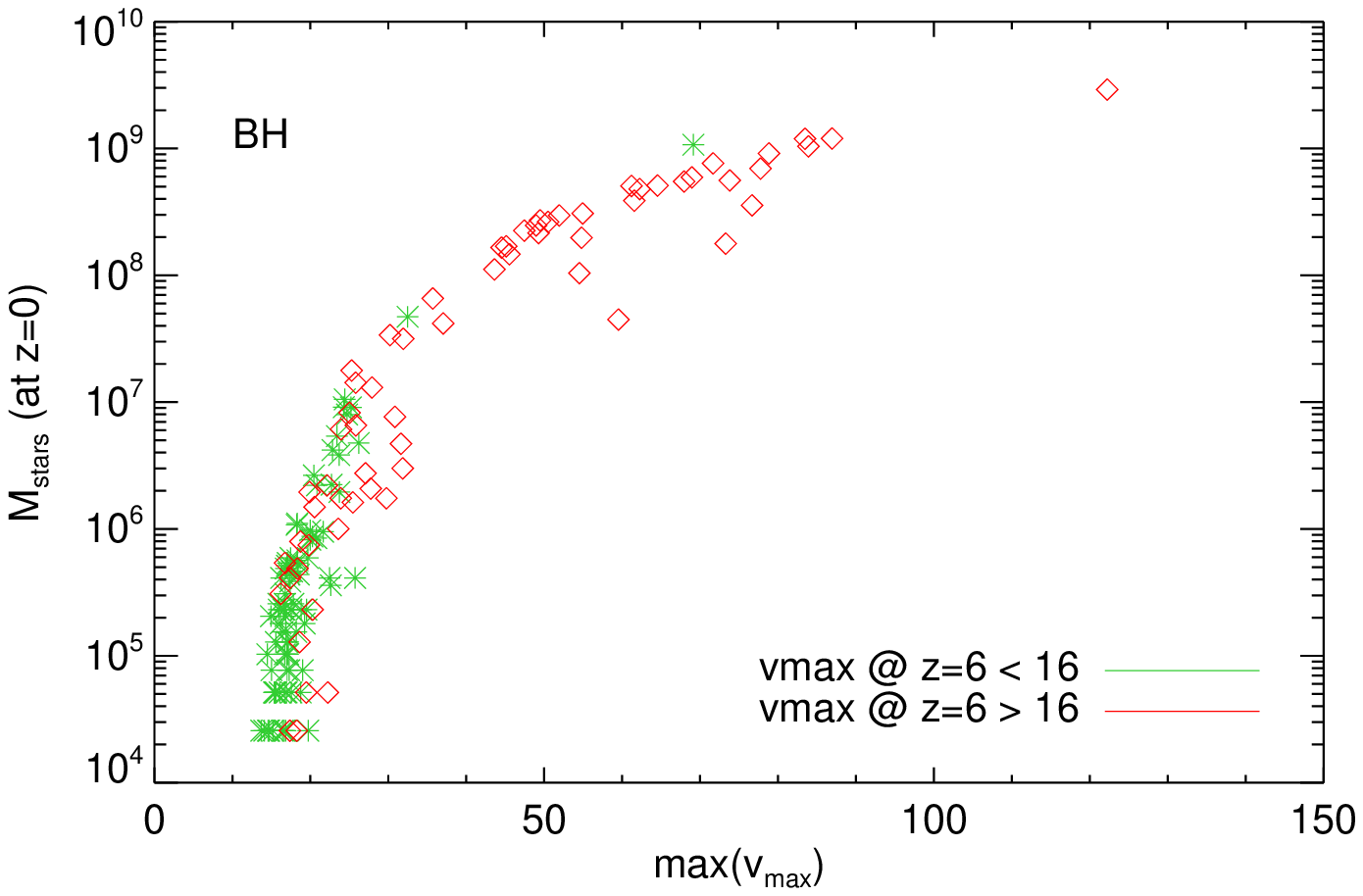} 
    \includegraphics[width=\linewidth]{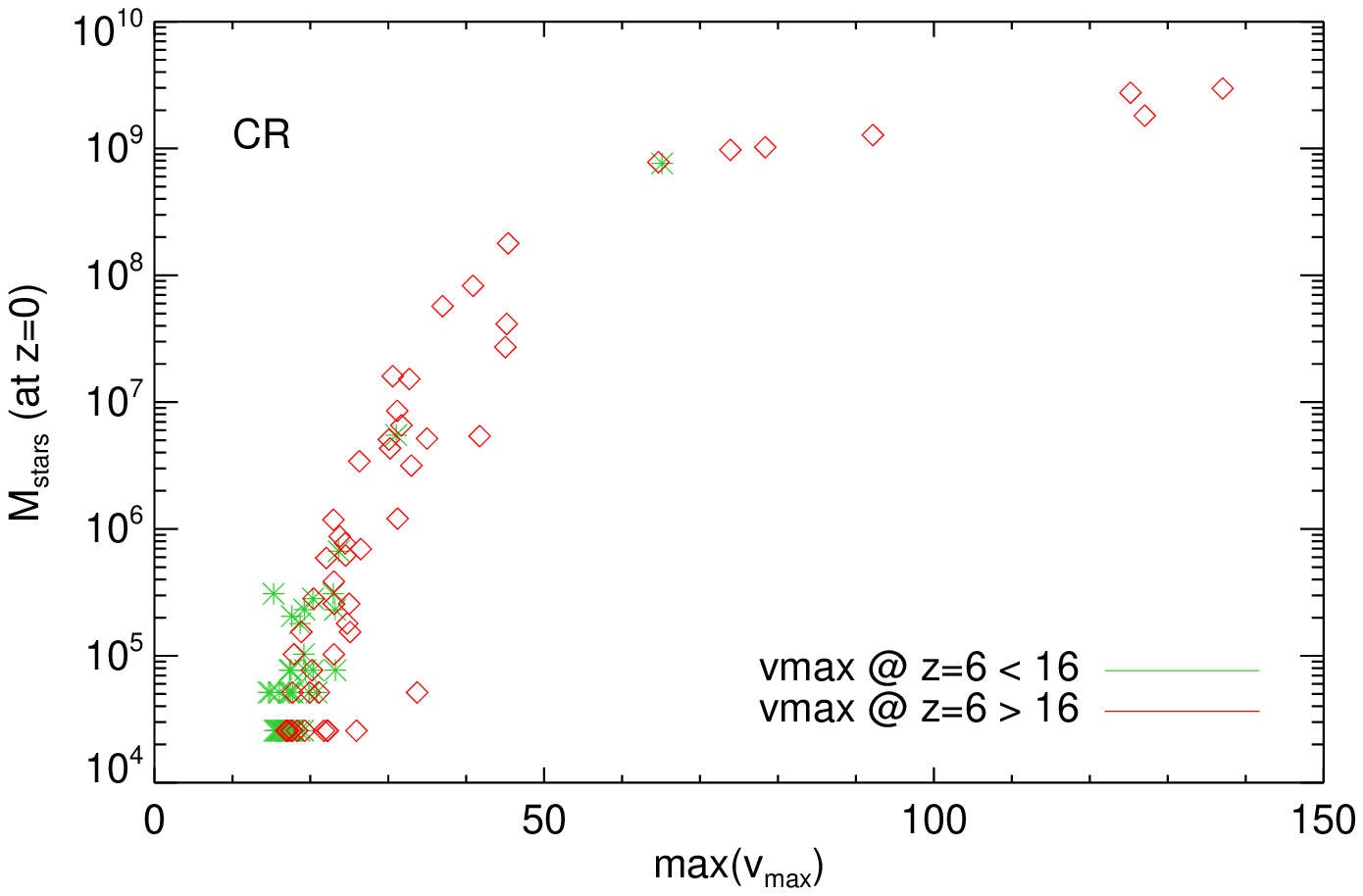} 
	\caption{Current stellar mass versus current maximum circular
          velocity. Satellite galaxies with $v_{\rm max}\ge 16 \;
          \mathrm{km/s}$ at $z = 6$ are shown as red diamonds, and
          satellites below this limit as green stars. There is again no
          evidence that this threshold or one nearby is able to
          distinguish between satellites with or without a significant
          amount of stars.
\label{fig:vmax_at_reio_cut}%
}%
\end{figure}

The different satellite histories we have selected in
Figs.~\ref{fig:Satellites_high_mass_bin}-\ref{fig:Satellites_low_mass_bin}
show a variety of interesting evolutionary effects that we now discuss
in turn. For definiteness, we have here selected the \textsc{BH} run, but our
other simulations show qualitatively very similar results.  The left
panel of Fig.~\ref{fig:Satellites_high_mass_bin} gives a nice
illustration of the tidal and ram pressure stripping effects that play
an important role in shaping the properties of the satellites. It is
clearly seen that the dark matter mass starts decreasing in distinctive
steps as soon as the satellite has fallen into the host halo and orbits
with rather high eccentricity. These mass stripping events correspond to
individual pericentric passages, as is clearly seen in the panel that
gives the distance to the host halo.  Note however that the stellar
component is not noticeably effected by this tidal stripping process, as
expected from the fact that the stars of the satellite are much more
concentrated than the dark matter.  In contrast, the gas component
behaves rather differently. Here we see clear evidence for ram pressure
stripping as the dominant source of mass loss even in high mass
satellites. Interestingly, this effect starts to set in even before the
satellite crosses the virial radius of the host, probably because the
gaseous halo of the host is more extended than $R_{200}$.

Because ram pressure stripping depends on the density of the surrounding
gas, one should expect to see variations of the mass loss rate with the
radial position of the infalling satellite. This is indeed seen if one
compares the results for the three different satellites shown in
Fig.~\ref{fig:Satellites_high_mass_bin} with each other. The satellite
on the left follows a very eccentric orbit and spends most of the time
in the outer parts of the halo, resulting in a comparatively slow
gaseous mass loss. The satellite shown in the middle panel has an orbit
with a lower eccentricity that keeps it at apocentre well inside the
virial radius, yielding a consistently higher mass loss rate. Finally,
the satellite shown on the right panel has a nearly circular orbit at
small radius, and loses its gas component even faster.

Perhaps one of the most interesting effects seen in Figures
\ref{fig:Satellites_high_mass_bin} to \ref{fig:Satellites_low_mass_bin}
is the effect of reionization on the star formation of the satellite
galaxies. Quite often the simplifying assumption is made that
reionization would be able to stop star formation in satellite galaxies
entirely, yet this is clearly in contradiction with the findings of our
simulations. In fact, all satellites shown in these Figures (and the
same is true for the majority of other satellites too) are producing
most of their stars at times later than $z=6$. Star formation continues
in all examples until the gas component is removed by ram pressure
stripping, but this time can be considerably later than the epoch of
reionization. We hence find that the detailed orbit of a satellite
galaxies tends to be more important for determining its final luminosity
than the circular velocity it had at the epoch of reionization. An
illustration of this can be seen in the histories of the satellites
shown in the middle and right panels of
Figure~\ref{fig:Satellites_medium_mass_bin}. These two satellites have
quite similar dark matter, gas and stellar masses shortly before they
enter the hot gaseous halo of the host galaxy, but they are moving on
very different orbits. The satellite in the middle panel is on a
relatively circular orbit, resulting in a small effect of tidal
stripping on the dark matter component and no noticeable effect on the
stellar component. In contrast, the satellite shown in the right panel
is on a highly eccentric orbit with $\epsilon \ge 10$. This satellite
dives deeply into the host halo, resulting in a tidal radius that is
even smaller than the characteristic radius of the stellar
component. Because of this, the stellar component looses nearly $90\;\%$
of its mass due to tidal effects.

In Figure~\ref{fig:vmax_at_reio}, we compare the final stellar mass of
the satellites against the maximum circular velocity $v_{\mathrm{max}}$
they had at redshift $z=6$. This provides another way to test the
popular hypothesis that the mass at the time of reionization determines
the final luminosity of a satellite galaxy. We show results for the
simulations \textsc{Ref}, \textsc{BH} and
\textsc{CR}\footnote{Unfortunately, all snapshots before $z=2.7$ of the
  simulation \textsc{Wind} where accidentally deleted, making this
  comparison impossible for this model.}. While most satellites
with circular velocities below $\sim 20\,{\rm km\,s^{-1}}$ are strongly
suppressed in stellar mass, there are a few objects with such low
circular velocities that have stellar masses as high as
$10^8\;\mathrm{M_\odot}$, or even $10^9\;\mathrm{M_\odot}$, at the present
epoch. In the range of $\sim 20\,{\rm km\,s^{-1}}$ to $\sim 30\,{\rm
  km\,s^{-1}}$, no sharp cut-off is readily apparent that could be
identified with reionization. Instead there is considerable scatter in
the relation between final stellar mass and maximum circular velocity at
$z=6$. We note that the simulation with cosmic rays shows a strong
suppression in the stellar mass for low circular velocities when
compared with the other simulations, as expected from our luminosity
function results.

A complementary of view of the above relation is shown in
Figure~\ref{fig:vmax_at_reio_cut}, where we plot the current maximum
circular velocity of satellites against their current stellar mass.
Different symbols encode the circular velocity they had at redshift
$z=6$, where satellites with $v_{\rm max} \ge 16 \; \mathrm{km\,s^{-1}}$ at
$z=6$ are shown as red diamonds while satellites below this threshold
are shown as green stars. Note that there is considerable overlap
between the regions occupied by the different symbols, showing again
that the correlation between the circular velocity at the epoch of
reionization and the current stellar mass is not overly strong.  In
particular, there are some examples (especially in the \textsc{BH}
simulation) of satellite galaxies with very low circular velocity at
$z=6$ which were nevertheless able to form many stars later on and to
turn into reasonably luminous satellites.

\begin{figure}
	\includegraphics[width = \linewidth]{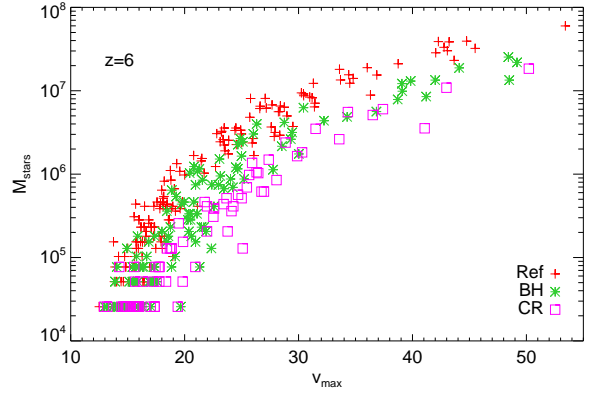}%
	\caption{Stellar mass versus maximum circular velocity at $z=6$,
          compared for different simulation models. Interestingly, a
          weak influence of AGN feedback on the satellite population is
          found at high redshift, but this difference largely vanishes
          later on, as most of the satellites are simply not able to
          grow a large supermassive black hole. In contrast, the
          additional pressure component of cosmic rays affects all
          satellites more strongly, and here the effect remains large in
          small galaxies even down to
          $z=0$.  \label{fig:Vmax_Mstar_at_reio}}
\end{figure}

Finally, in Figure~\ref{fig:Vmax_Mstar_at_reio}, we directly compare the
results of the different physics simulations with each other, in terms
of the relationship between $v_{\max}$ and the stellar mass at the time
of reionization. Interestingly, we see that AGN feedback is in fact able
to reduce the stellar mass formed at high redshift in many of the
progenitor systems of today's satellites, even though the effect is
considerably weaker than for cosmic rays. As we have already seen, the
influence of BH feedback tends to become however low at later times, so
that the present day properties of satellites are only mildly
effected. This is presumably because most satellites do simply not grow
a massive black hole, but they are nevertheless affected at high
redshift by the seed black hole that is injected into their halo.

\begin{figure*}
	\centering
  \begin{minipage}[t]{0.47\textwidth}
    \centering
    \includegraphics[width=\textwidth]{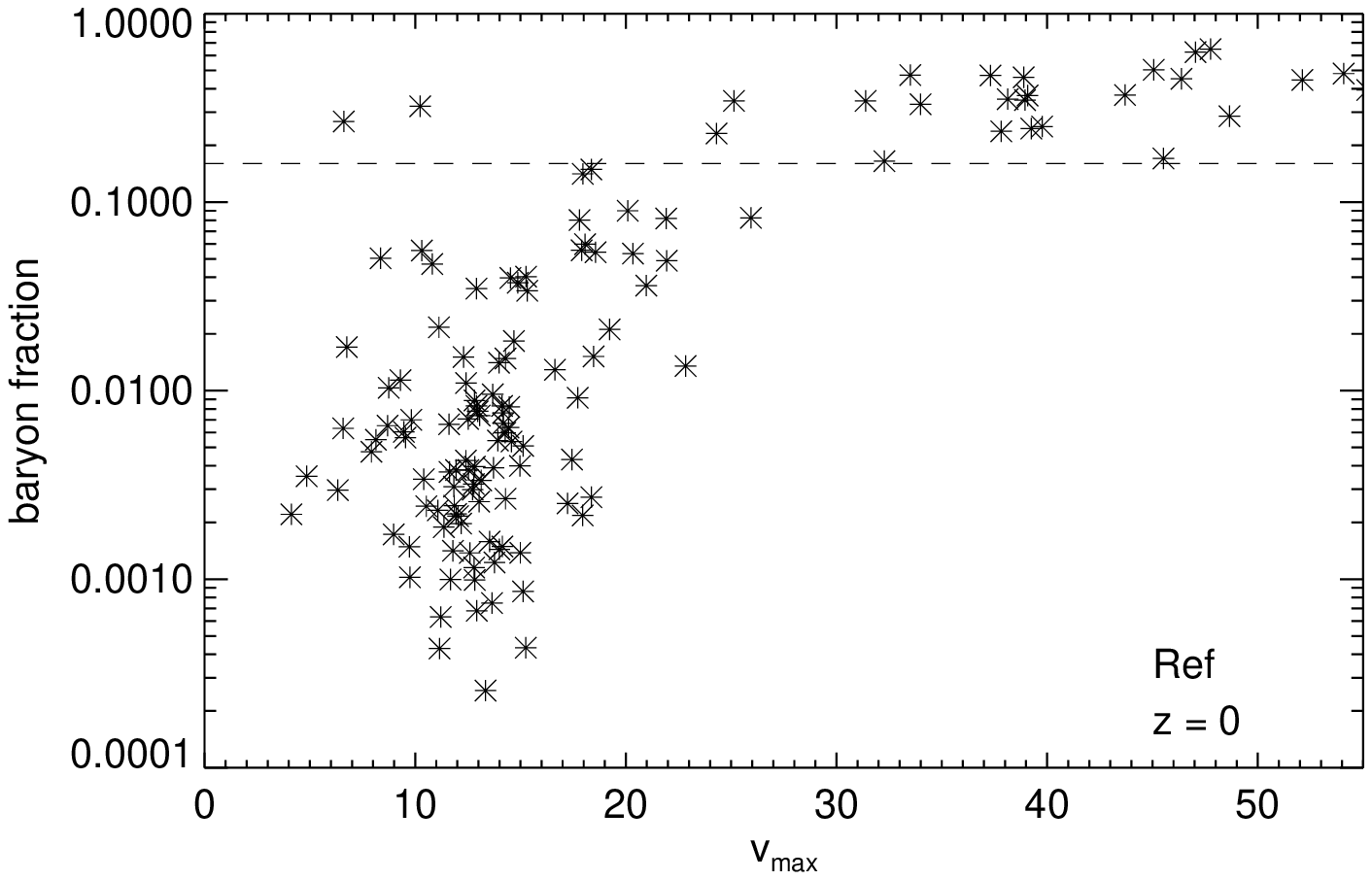} 
  \end{minipage}
  \begin{minipage}[t]{0.47\textwidth}
    \centering
    \includegraphics[width=\textwidth]{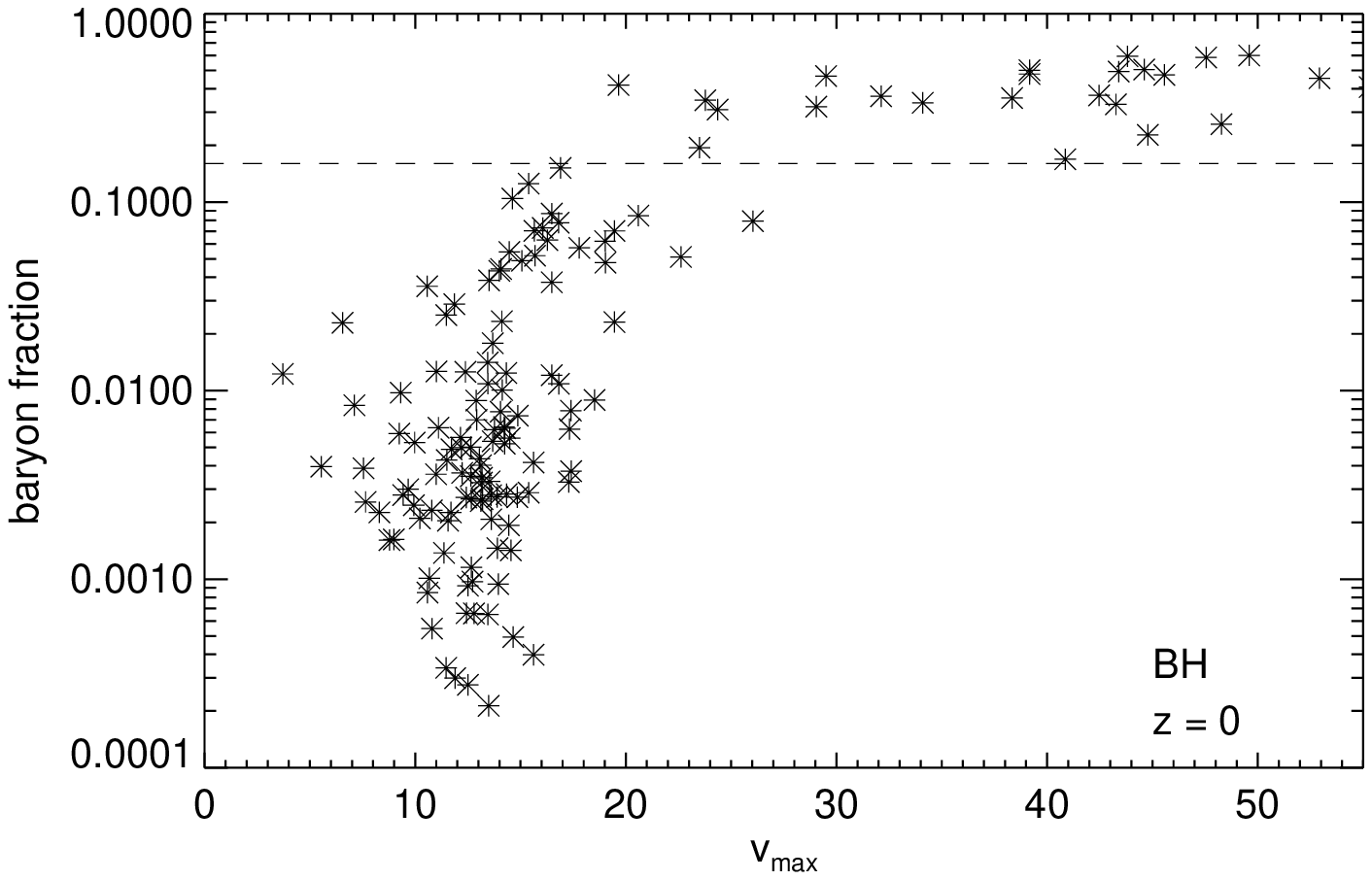} 
  \end{minipage}
  \hfill
  \begin{minipage}[t]{0.47\textwidth}
    \centering
    \includegraphics[width=\textwidth]{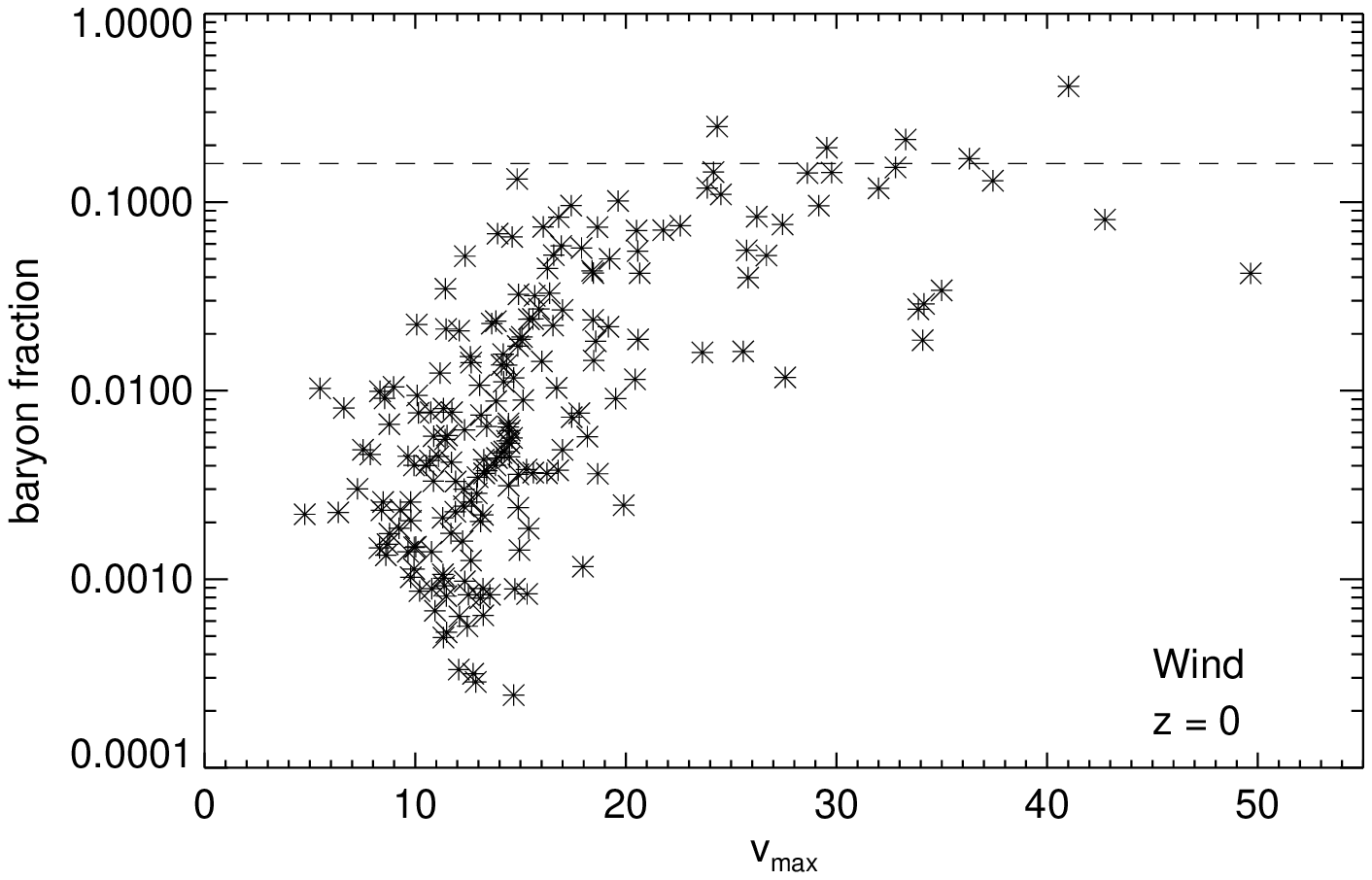} 
  \end{minipage}
  \begin{minipage}[t]{0.47\textwidth}
    \centering
    \includegraphics[width=\textwidth]{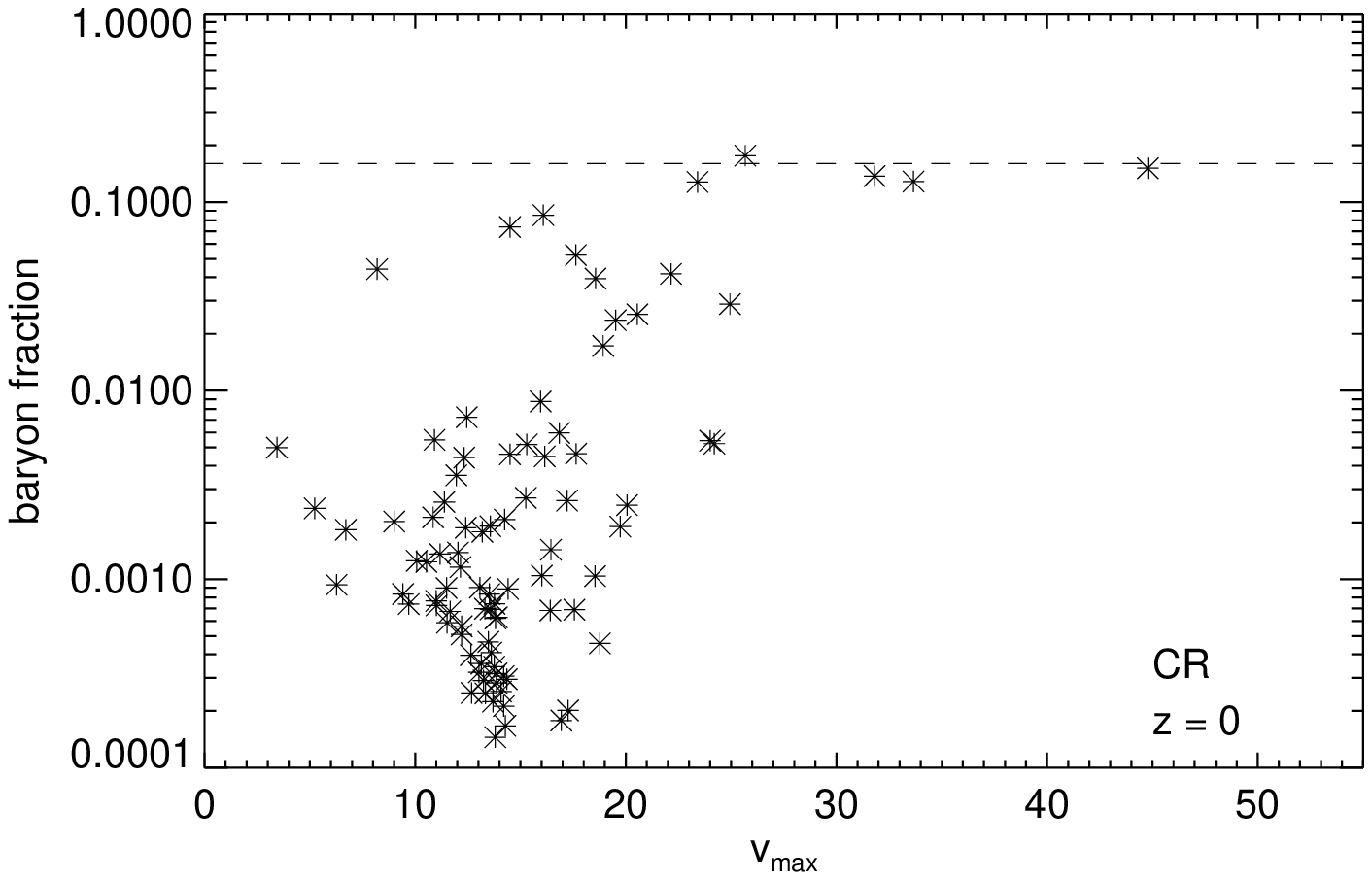} 
  \end{minipage}
  \hfill
	\caption{Baryon fraction of $z=0$ satellite galaxies as a
          function of their circular velocity. We show results for our
          four primary simulation models, and define the baryon fraction
          in terms of the bound particles identified by {\small SUBFIND}
          for each satellite. The horizontal dashed line gives the
          universal cosmic baryon fraction. \label{fig:BarFrac_now}}
\end{figure*}

In Figure~\ref{fig:BarFrac_now}, we plot the baryon fraction of
satellites against $v_{\max}$, at the present epoch.  The baryon
fraction is here simply defined as the total bound baryonic mass
relative to the total bound mass of a halo. It is interesting to compare
this value with the universal cosmic baryon fraction
$\Omega_b/(\Omega_b+\Omega_{\rm dm}) = 0.16$ (shown as dashed horizontal
line).  As one expects, the baryon fraction is usually lower than the
cosmic baryon fraction, especially for very low mass satellites that
have lost most of their gas and did not form many stars either.
However, in the simulation with comparatively weak feedback, some
satellites have also baryon fractions above the cosmic mean value. These
are satellites which lost a lot of dark matter through tidal stripping
whereas they could hold on to most of their stars. Both the
\textsc{Wind} and \textsc{CR} models are leading to considerably reduced
baryon fractions in low mass satellites. In the former case, this is
readily expected as a signature of the winds. In the latter, it is
because more baryons stay in a diffuse gaseous phase, allowing them to
be more easily ram-pressure stripped.

\begin{figure}
    \includegraphics[width=\linewidth]{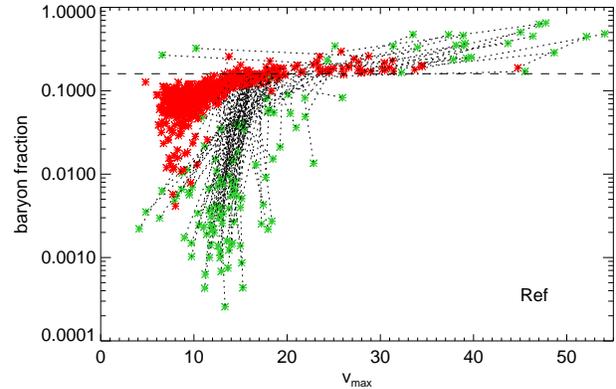} 
	\caption{Evolution of the baryon fraction and circular velocity
          between $z=6$ and $z=0$. For each satellite, we mark the high
          redshift $z=6$ data with a red star and the $z=0$ value with a
          green star, and we connect each pair of two points with a
          dotted line. There is clearly a pivotal maximum circular
          velocity of about $\sim 25\,{\rm km\,s^{-1}}$ below which most
          satellites lose a large fraction of their baryons, while they
          retain their baryon fraction above this threshold.
	\label{fig:BarFrac_evo}}
\end{figure}

An analysis of the evolution of the baryon fraction between $z=6$ and
$z=0$ is given in Figure~\ref{fig:BarFrac_evo}. We here only show
results for the \textsc{Ref} simulation, as the qualitative behavior of
the other simulations is similar.  We use two symbols for each
satellite, one showing the data point at $z=6$ (red stars), while the
corresponding values at $z=0$ are given by green stars. Every pair of
points belonging to the same satellite is connected by a dotted line.
Most satellites with circular velocities below $\sim 20\,{\rm km\,
  s^{-1}}$ at $z=6$ lower their baryon fraction substantially until the
present epoch, and they also do not tend to grow much. In contrast, most
larger satellites tend to keep their baryon fraction or increase it
slightly, often accompanied by a significant increase in $v_{\max}$.  In
the most massive satellites, part of this increase stems from
modifications of the inner rotation curve due to the formation of a
quite concentrated stellar component, i.e.~these satellites are not
really bona-fide dark matter dominated systems as often assumed.  We
note however that the threshold at $\sim 20\,{\rm km\, s^{-1}}$ is not
sharp; there are still many examples of satellites with an initially
high $v_{\max}$ that end up as low mass satellites with a stripped
baryonic component.

\begin{figure*}
	\centering
  \begin{minipage}[t]{0.47\textwidth}
    \centering
    \includegraphics[width=\textwidth]{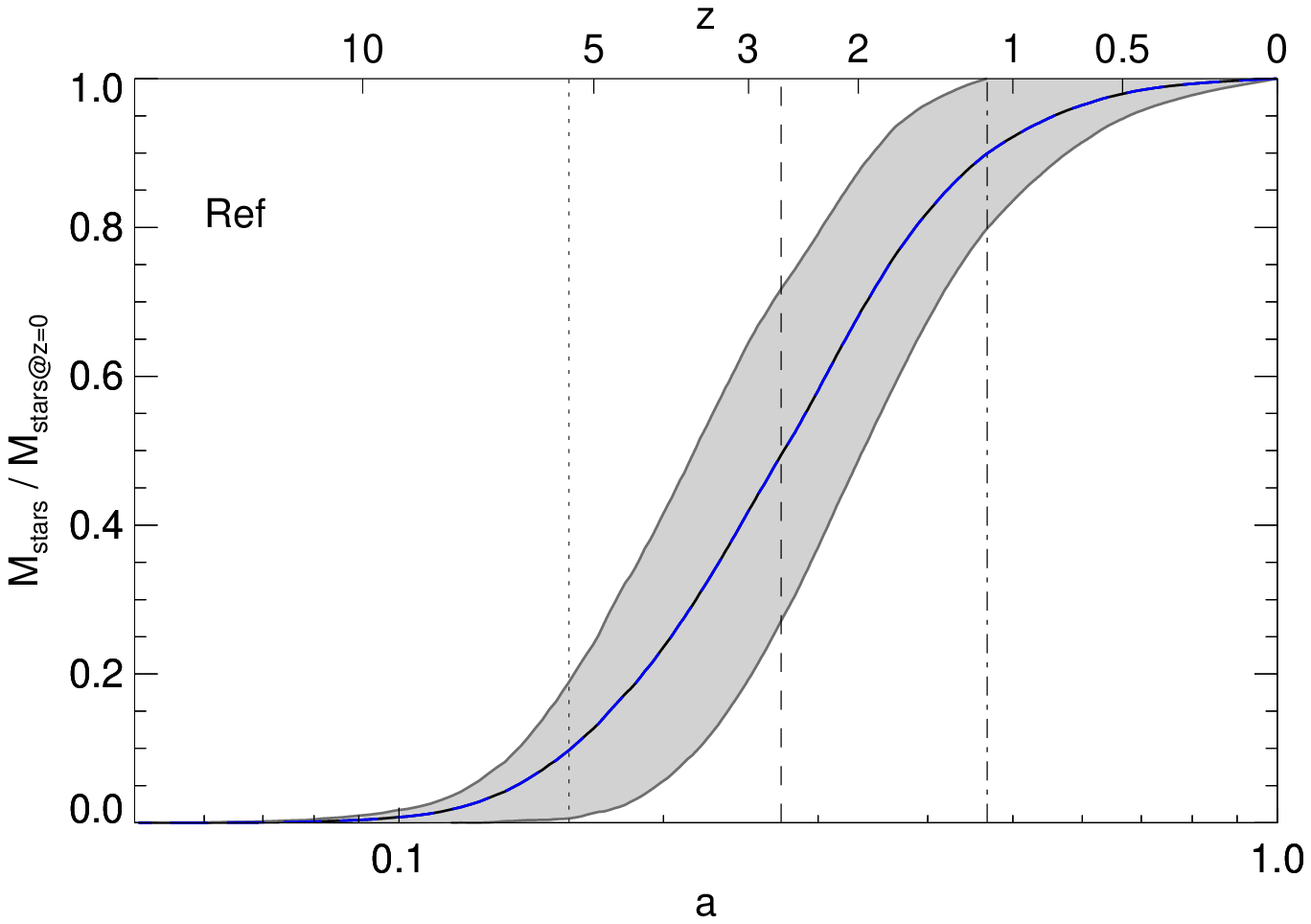} 
  \end{minipage}
  \begin{minipage}[t]{0.47\textwidth}
    \centering
    \includegraphics[width=\textwidth]{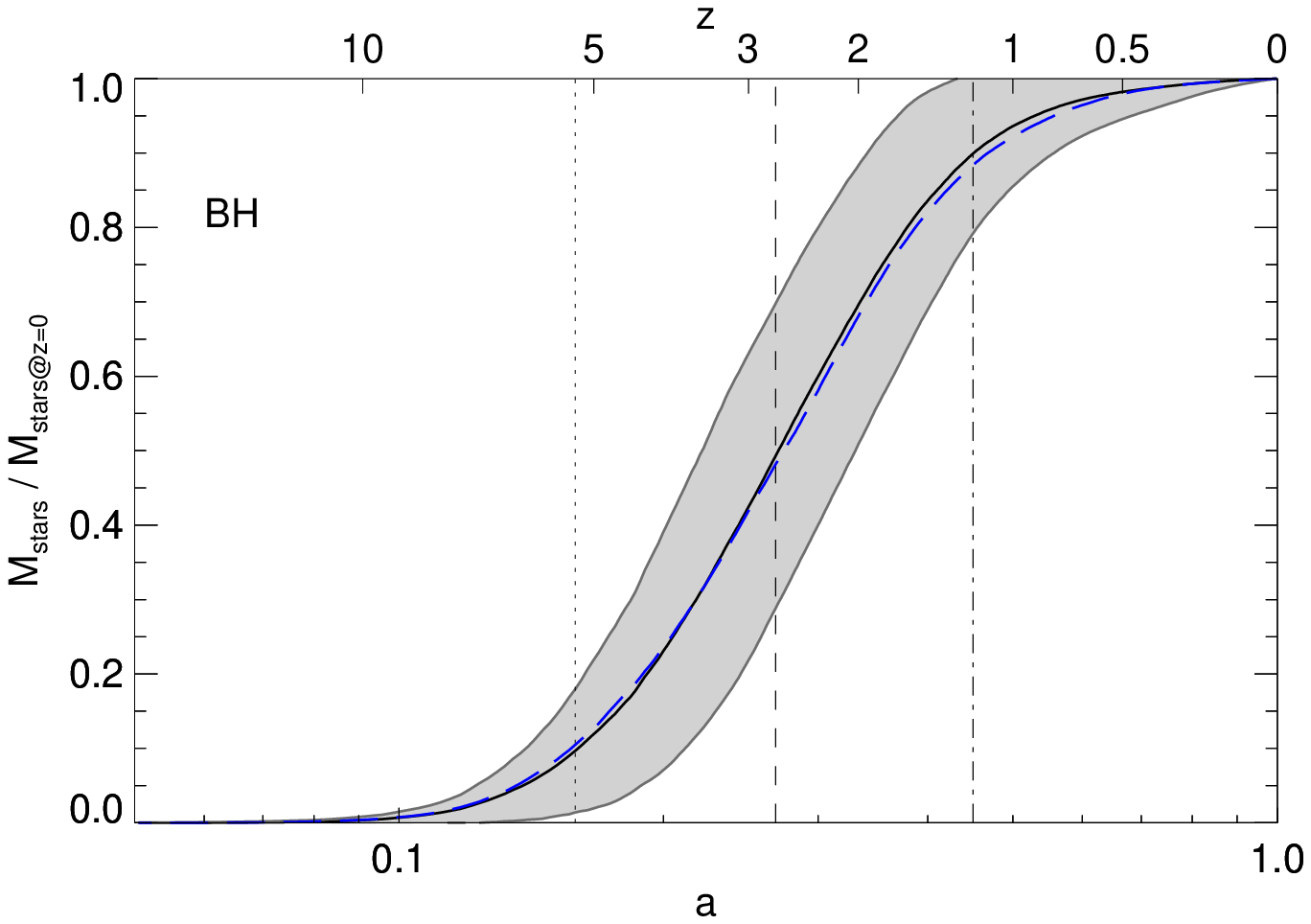} 
  \end{minipage}
  \hfill
  \begin{minipage}[t]{0.47\textwidth}
    \centering
    \includegraphics[width=\textwidth]{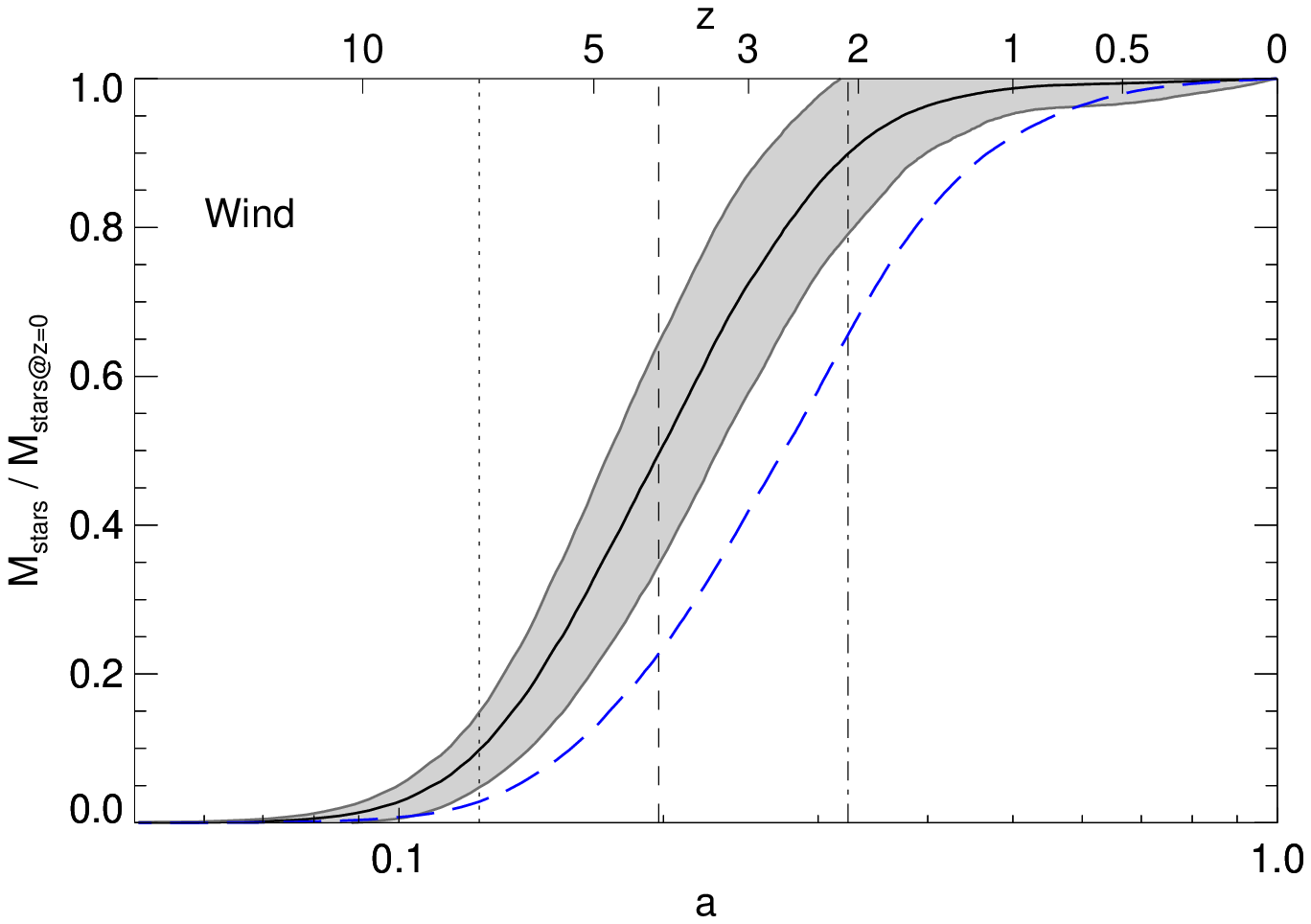} 
	\hfill
  \end{minipage}
  \begin{minipage}[t]{0.47\textwidth}
    \centering
    \includegraphics[width=\textwidth]{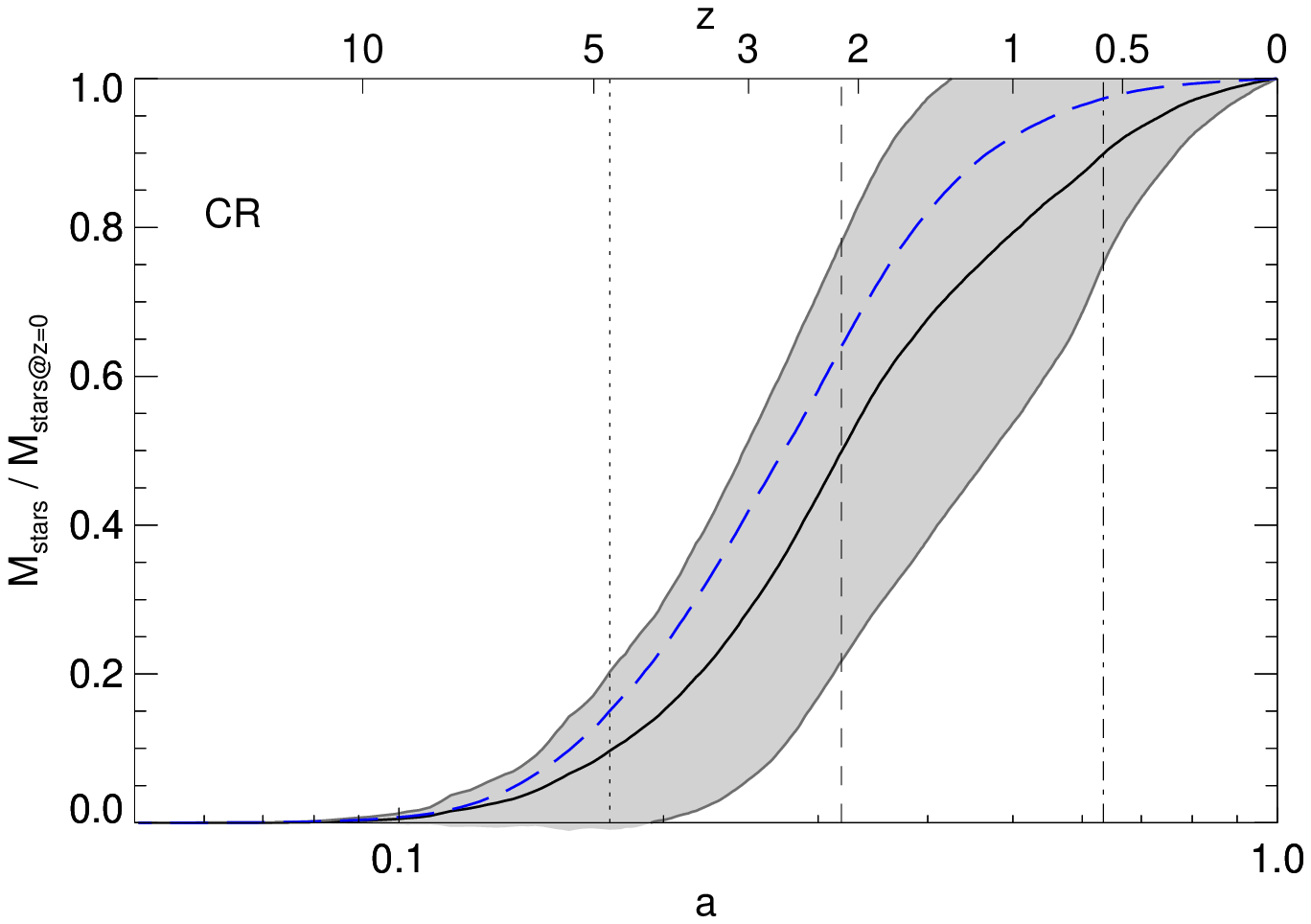} 
  \end{minipage}
  \hfill
	\caption{Average cumulative star formation histories of all
          satellite galaxies, for all four primary simulation models.
          In each panel, the solid line shows the average star formation
          history, with the grey bands mark the $1 \; \sigma$ scatter
          of the distribution. The vertical dashed and dotted lines give
          the times when $10\%$, $50\%$, and $90\%$ of all the stars have formed.
          To ease the comparison between the different simulations, the
          result of the \textsc{Ref} simulation is repeated in all the
          panels as a dashed blue line.
	\label{fig:Satellites_SFRHistory}}
\end{figure*}

The last quantity we analyze in this section are the cumulative star
formation histories of our satellites, as shown in
Figure~\ref{fig:Satellites_SFRHistory}.  The solid black line shows the
total cumulative star formation history of all satellites in the final
virial radius, normalized by their total final stellar mass. The gray
shaded area gives the $1\sigma$ scatter around this mean for the
ensemble of all satellite star formation histories.  The vertical
dotted, dashed and dot dashed lines mark the times when $10\,\%$,
$50\,\%$ and $90\,\%$ of the stars present at $z=0$ were
formed. Finally, the dashed blue line repeats the result of the
\textsc{Ref} simulation in all the panels corresponding to the other
simulations, in order to ease a comparison between them. As we have
already seen in the other results, AGN feedback shows little effect on
the cumulative star formation history of the satellites. The
\textsc{Wind} model on the other hand leads on an earlier production of
the bulk of the stars, which is what one would expect if galactic
outflows are efficiently removing gas from star-forming dwarf galaxies
and are thus shutting down star formation earlier. In contrast, the
\textsc{CR} model shows exactly the opposite effect. Due to the
additional pressure component, the galactic gas has a lower overall
cooling rate. This hampers star formation in low mass systems but does
not by itself remove significant amounts of fuel for star formation; the
latter can however be achieved by ram pressure stripping. Thus, star
formation shifts to considerably later times in the \textsc{CR} run than
in any of the other models.

Interestingly, the scatter around the mean history of the satellites is
also modified by the different physics. The \textsc{Wind} simulation
shows a rather small scatter, presumably because most satellites form
their stars in the first significant phase of star formation at high
redshift, which is terminated quickly and for the most part coevally.
In the case of the \textsc{CR} simulation, much of the gas is not
removed by the primary feedback process itself, but instead is affected
by stripping processes at intermediate and low redshifts, after the
satellites have fallen into the parent halo. This means that the
individual infall history of each satellite is of larger importance in
this model, leading to a higher overall variability in the star
formation history of the satellites.

\section{Scaling relations}\label{sec:scaling}

In this section, we investigate in more detail how the properties of our
simulated satellites scale with their size. Where possible, we compare
with observational results and other theoretical predictions.  We want
to caution however that especially our smallest luminous satellites are
pretty close to our resolution limit.  While the satellites above the
detection limits of SDSS should be sufficiently well resolved in our
high resolution simulation to give reliable results, a considerable
numerical uncertainty persists, a fact that should be taken into account
in interpreting the results.

\begin{figure*}
	\centering
\includegraphics[width = 0.9\linewidth]{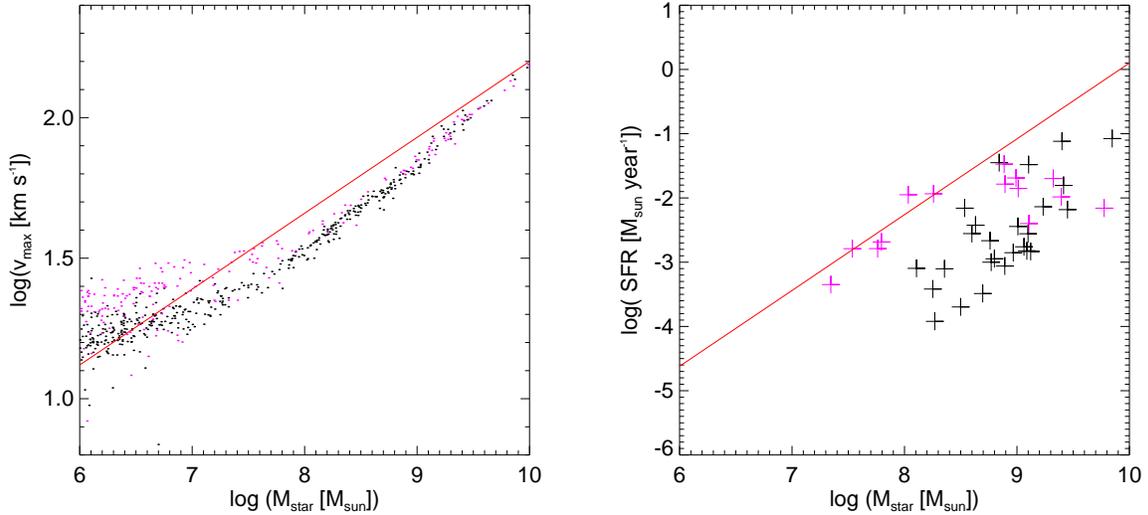}%
	\caption{The left panel shows the relation between stellar mass
          and maximum circular velocity at the present time, while the
          right panel gives the relation between stellar mass and star
          formation rate. The red lines give the best fits
          \citet{DwarfScalingRelations} derived for the observational
          data. We here included all simulated dwarf galaxies within the
          full high resolution region of the \textsc{Ref} simulation (in
          black) and of the \textsc{CR} simulation (in magenta), since
          \citet{DwarfScalingRelations} did also include dwarfs outside
          of the virial radius of the Milky Way.
	\label{fig:Satellites_Scaling_Relations}}
\end{figure*}

We begin with the scaling relations derived by
\citet{DwarfScalingRelations} for local group dwarf galaxies.  We focus
on the relations between stellar mass and circular velocity, and stellar
mass and star formation rate, as they have the highest statistical
significance and are thus best suited to benchmark the simulation
results. Figure~\ref{fig:Satellites_Scaling_Relations} shows these two
relations in separate panels, comparing in each case the fits derived by
\citet{DwarfScalingRelations} with our simulation data.  The correlation
between stellar mass and maximum circular velocity is comparatively
tight, in fact, \citet{DwarfScalingRelations} cite a correlation
coefficient of $0.94$ for the observations, which are represented by the
solid red line. The simulated satellites show a similarly strong
correlation (formally yielding a correlation coefficient of $0.94$), but
the results for the \textsc{Ref} simulation are slightly offset towards
higher stellar masses. The \textsc{CR} results (shown with magenta
symbols) agree considerably better.  Our lowest mass satellites start to
deviate from the fit given by \citet{DwarfScalingRelations} but the
differences are of comparable size as in some observed systems such as
Ursua Minor, and the growing scatter in the numerical results also
indicates that resolution effects start to play a role.

The right panel in Fig.~\ref{fig:Satellites_Scaling_Relations} compares
the correlations between stellar mass and current star formation
rate. Here we find a much worse agreement with the observational results
of \citet{DwarfScalingRelations}, which are again characterized by a
remarkably good correlation (with coefficient $0.96$). While our results
bracket the observationally inferred relation, the scatter is large and
the formal correlation is only $0.72$. In addition, most of our
satellite galaxies show only a vanishingly small star formation rate at
the present epoch, and those systems were omitted from the plot. But again, 
the \textsc{CR} results seem to agree better with observations.
Simulations with better mass resolution will be needed to shine more
light on this potential discrepancy.

\citet{SatKinematics} calculated the mass-to-light ratio for the sample
of Milky Way satellites found in the SDSS, obtaining values ranging from
about $100$ to $1200$, with a mean of $\sim 380$. Doing the same
calculation for the whole sample of known satellites resulted in values
between $1.5$ and $1200$ with a mean of $\sim 170$. This suggests that
the faint satellites discovered with the SDSS are even more dark matter
dominated than the more luminous `classical' satellites.  For the
complete sample of simulated luminous satellites, we obtain
mass-to-light ratios between $12 \;(11, 16, 10)$ and $13000 \;(18000,
23000, 18000)$ with a mean of $\sim 1500 \;(1350, 1550, 3500)$ for the
\textsc{Ref} (\textsc{BH}, \textsc{Wind}, \textsc{CR}) simulation. As
mentioned earlier, very small satellites are strongly affected by
numerical effects and thus the very large mass-to-light ratios we find
for these satellites may be unreliable. Also, the mean value may be
biased high by the large number of small satellites. If we restrict the
mass-to-light ratio calculation to satellites with $M_V \le -8.0$,
we obtain a mean of $232 \;(217, 315, 345)$, which is much closer to
the observed values. If we select only satellites with a surface
brightness $\mu \le 30$, then the mean mass-to-light ratio is $33 \;(33,
73, 33)$, which is about $5$ times smaller than the observed mean value
for this selection.

\begin{figure}
	\centering%
\includegraphics[width =  \linewidth]{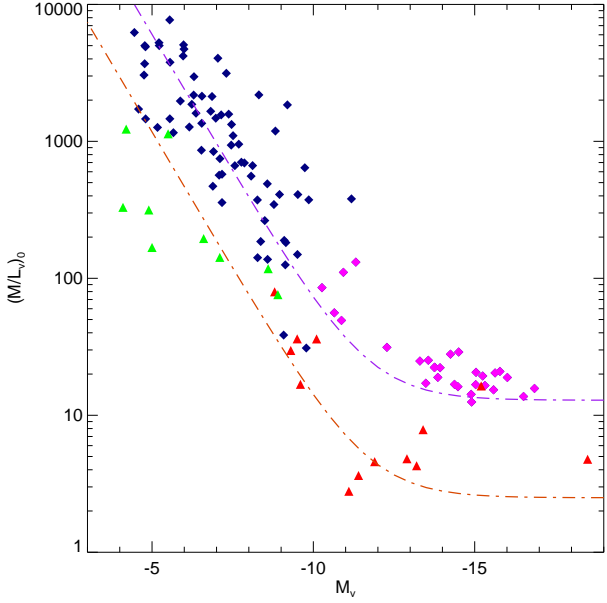}%
	\caption{The relation between mass-to-light ratio and luminosity
          of satellite galaxies \citep[see][]{SatData2}. The observed
          satellites are plotted as triangles, with red symbols marking
          satellites known before SDSS and green symbols marking the
          newly discovered satellites discovered in the SDSS
          data. Diamonds give our simulated satellites (\textsc{Ref}
          model), colour-coded as magenta if their surface brightness is
          high ($\mu \ge 30$) or as dark blue if it is low. The
          dot-dashed lines represent the fitting function suggested by
          \citet{SatData2}, for the observed sample (red) and shifted
          upwards by a factor of $5$ (dark purple) to match the
          simulated sample.}%
	\label{fig:Mateo_Plot}
\end{figure}

This difference in the mean mass-to-light ratio can also be seen from
the ``Mateo Plot'' shown in Figure~\ref{fig:Mateo_Plot}, which compares
the luminosities of the satellite galaxies with their mass-to-light
ratio in units of the solar mass-to-light ratio. In the original paper
where this plot was introduced,
 \citet{SatData2} overplotted the function
\begin{equation}
	\log\frac{(M/L)}{(M/L)_\odot} = 2.5 + 10^7/(L/L_\odot),
\end{equation}
which we also included as the dark red dot-dashed line in
Fig.~\ref{fig:Mateo_Plot}. To guide the eye, we also simply scaled this
function by a factor of $\approx 5.2$ and plotted it again as the purple
dot-dashed line.  It can be seen that this scaled function fits the
simulated galaxies of the \textsc{Ref} simulation very well, while the
observed systems (shown with red and green triangles) are well described
by the original function. This result is consitently reproduced by all
simulations.  Only the constant horizontal offset between observations
and simulations changes by $\approx\;10\%$ while the general trend
remains the same.

We note that the nearly constant offset between the simulated and
observed satellite galaxies could in part be caused by the rather
uncertain procedure applied to estimate the total mass of observed
satellites. This effectively involves an extrapolation to the outer edge
of the satellite, which is uncertain.  An alternative would be that the
simulated galaxies simply contain fewer stars than expected for an
observed satellite of the same mass.  However, the \textsc{Ref}
simulation already has comparatively weak feedback, and allowing for
brighter satellites by a constant factor would cause the most luminous
satellites, which are in good agreement with the Magellanic Clouds, to
become too bright.  Furthermore, making the star formation more
efficient in all satellites would shift the points in
Fig.~\ref{fig:Mateo_Plot} both down and to the right, hence spoiling the
good agreement with the location of the break in the observed relation.

\begin{figure}
	\centering%
	\includegraphics[width = \linewidth]{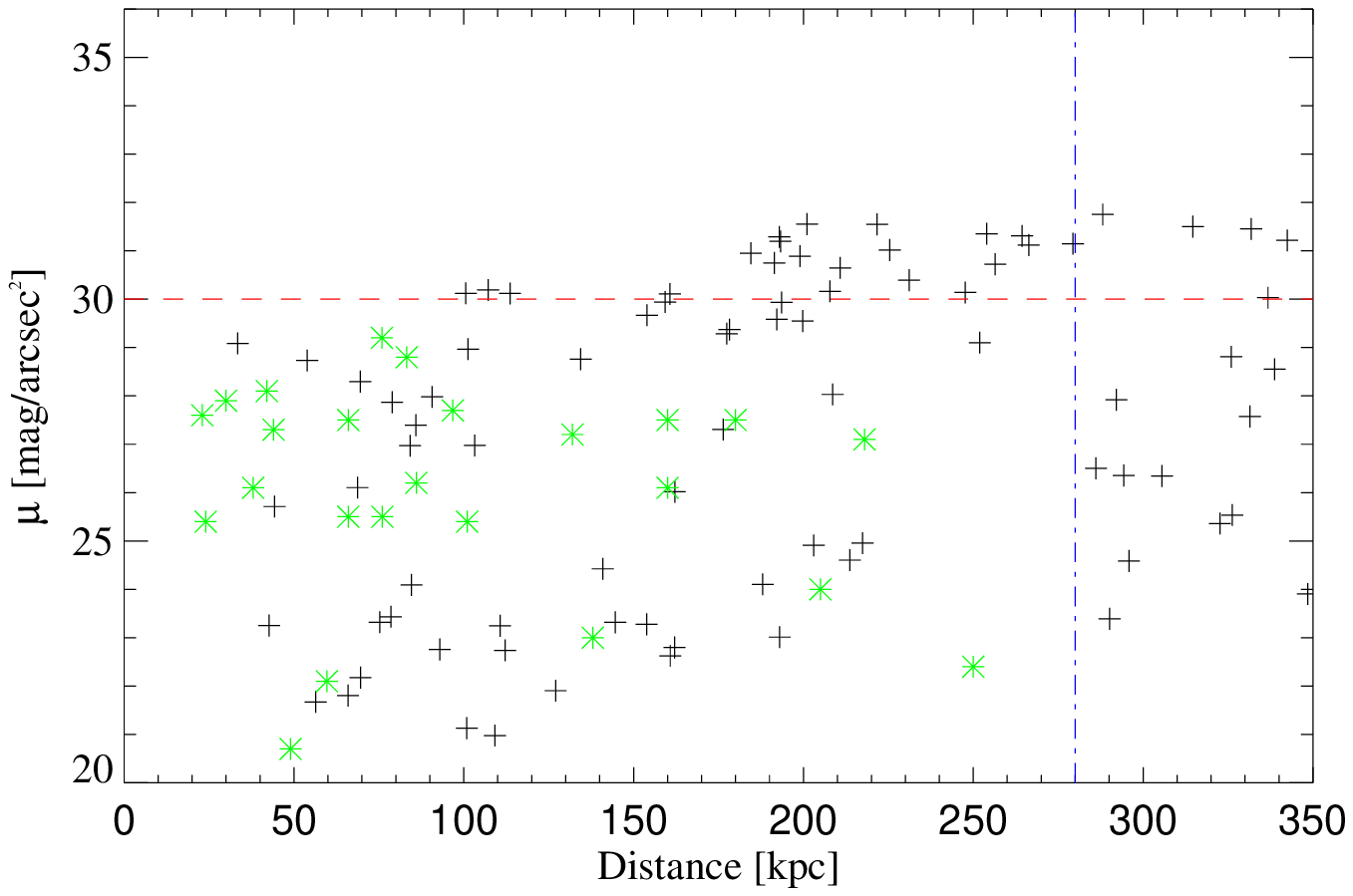}%
	\caption{Photometric V-band surface brightness of the simulated
          satellites (black crosses), inside a sphere of radius $350\,
          \mathrm{kpc}$ around a fiducial position of the Sun in the
          simulations. The red line shows the surface brightness
          detection limit of the SDSS survey, while the blue vertical
          line gives the radial cut at $280 \, \mathrm{kpc}$ that we
          frequently used in this work. There are $46$ Satellites below
          this limit within a radius of $280\, \mathrm{kpc}$ which is
          relatively close to the $57$ satellites predicted for an
          all-sky extrapolation of the observational data. The green
          stars give the observed satellites with well determined
          surface brightnesses.
	\label{fig:Satellites_Vmag_phot}}
\end{figure}

In Figure~\ref{fig:Satellites_Vmag_phot}, we show the relation between
photometric surface brightness of all simulated dwarf galaxies inside a
sphere of radius $350\; \mathrm{kpc}$ centred on a fiducial position of
the Sun. The Sun was assumed to lie $8.5\; \mathrm{kpc}$ away from the
centre of the galaxy, in the central plane of the stellar disc. As can
be seen from the relatively large scatter of the plot, the simulation
produces also satellites that are well above the SDSS surface brightness
detection limit. Counting the galaxies with a photometric magnitude
brighter then $30\; \mathrm{mag/arcsec^2}$ (dashed red line) and a
distance smaller than $280\; \mathrm{kpc}$ (dot dashed blue line)
results in observable $46$ ($77$, $18$, $70$) satellites for the
\textsc{Ref} (\textsc{BH}, \textsc{CR}, \textsc{Wind}) simulation. This
is actually in reasonable agreement with the prediction of $57$
satellites for the Milky Way.  The cutoff radius of $280\; \mathrm{kpc}$
was chosen as a compromise between the measured distances to all known
satellites, which reach up to $\approx 1\;\mathrm{Mpc}$, and the virial
radius of $r_{200}= 238\;\mathrm{kpc}$ of the simulated host galaxy.
This is also the same cut off radius that has been used in previous work
\citep{Koposov08,Maccio10}, although we note that some studies have
adopted a different choice \citep[e.g.][]{Via_Lactea_I}. The rather
small number of satellites classified as `observable' for the
\textsc{CR} simulation can easily be explained by the lower luminosity
function shown in figure \ref{fig:Satellites_Luminosity_Function}.

\begin{figure}
	\centering
    \centering
    \includegraphics[width=\linewidth]{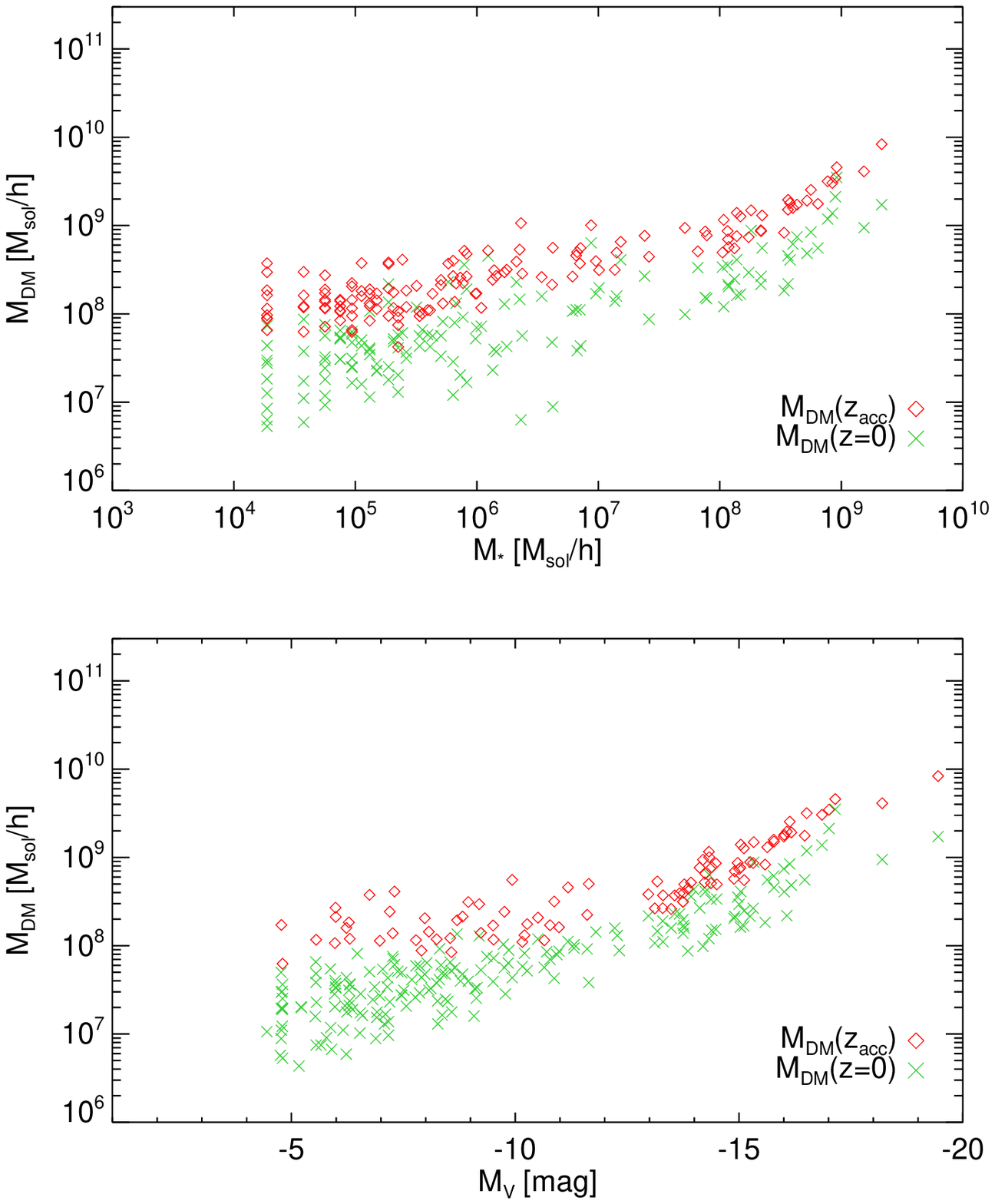} 
	\caption{Relation between dark matter satellite mass and stellar
          mass, or stellar luminosity, respectively. For each satellite
          we show both the dark matter mass today and at redshift
          $z=6$. This plot may be compared directly to figure 5 of
          \citet{Maccio10}, which is representative for semi-analytic
          models constructed to describe the satellite
          population. Unlike in their results, there is clearly some
          curvature found with the hydrodynamical simulation in this
          relation.
	\label{fig:maccio_5}}
\end{figure}

Finally, we consider the relation between the dark matter masses of our
simulated satellites with their stellar mass and luminosity, as shown in
Figure~\ref{fig:maccio_5}. For each satellite, we plot the dark matter
mass with different symbols, both at the epoch of accretion and at the
present epoch.  To simplify a comparison with figure 5 of
\citet{Maccio10}, we used exactly the same axis range in our plot as
they did.  Unlike in the results of \citet{Maccio10}, we find a clear
bend in the relation, meaning that our satellites tend to have higher
stellar masses, especially at the low mass end, than the satellites of
\citet{Maccio10}.  The latter results are based on a semi-analytic model
where the orbits of an infalling satellite are estimated based on a
random choice of plausible infall parameters.  It is possible that this
explains the discrepancy, or that it originates in approximate
treatments of tidal or ram pressure stripping in the semi-analytic
model. In future work, it will be interesting to inter-compare direct
hydrodynamical simulations and the semi-analytic models on a satellite
by satellite basis, in order to better understands the origin of these
differences in the predictions.

\section{Conclusions}\label{sec:conclusions}

In this work, we studied a set of high-resolution hydrodynamical
simulations of the formation of a Milky Way sized galaxy, starting from
cosmological initial conditions. Such simulations are now able to reach
sufficiently high resolution to directly resolve the formation of the
small dwarf galaxies that orbit in the halo, thereby allowing studies of
the missing satellite problem and of the properties predicted by
simulations for the population of satellite galaxies. These galaxies are
especially interesting both because the dark matter substructure
abundance is a fundamental challenge for the $\Lambda$CDM cosmology, and
because the low star formation efficiencies of the satellites provide
crucial information about the physics of feedback.

We have therefore repeated our simulations using different models for
feedback physics, with the goal to test the sensitive of the results for
the satellites with respect to these physics assumptions.  In the
\textsc{Ref} model, we considered only star formation and SN feedback,
together with instantaneous reionization at $z=6$. The three other
models included additional processes like AGN feedback (\textsc{BH}),
wind driven galactic outflows (\textsc{Wind}) and the generation and
decay of cosmic rays (\textsc{CR}). Not unexpectedly, the \textsc{BH}
model showed no significant differences compared to the reference
\textsc{Ref} model, as most of the satellites are simply too small to
grow a large supermassive black hole and are rarely affected by strong
quasar feedback in neighbouring galaxies. In contrast, the \textsc{Wind}
model showed a significant reduction of the number of high mass
satellites, but did not give a significantly different abundance of low
mass systems.  The \textsc{CR} model had exactly the opposite effect as
it did not change the high mass satellites but suppressed star formation
in low mass satellites. This made the cosmic ray model most successful
in matching the faint-end of the observed satellite luminosity
function. Our results further suggest that a combination of the
\textsc{Wind} and \textsc{CR} feedback models should be able to yield a
nearly perfect match of the luminosity function.

The total number of satellites observable with an SDSS-like survey
covering the whole sky has been estimated to be $57$
\citep{SatKinematics}. Interestingly, imposing the same surface
brightness detection threshold on all of our simulated systems yields a
prediction of $77$ observable satellites for our \textsc{BH} model,
which is only moderately higher than the observations despite the fact
that this simulation overpredicts the satellite luminosity function
considerably. For our \textsc{CR} model instead, the number drops
considerably, to $18$, perhaps caused in part by an overprediction of
the effective stellar radii of the satellites, which could easily arise
from the limited spatial resolution of our simulations. In any case,
this stresses that a large number of additional satellites may actually
still be hidden just below the surface brightness limit of the SDSS
\citep[see also][]{Bullock2009}.

Our simulations have also highlighted the relative importance of some of
the evolutionary aspects of satellite galaxies. In particular, we do not
find a very distinctive mark of the epoch of reionization on the
satellites, and most satellites continue their star formation activity
in our simulations to much lower redshift than $z=6$. This suggests that
simplified treatments of satellite histories, where relatively high
cooling thresholds due to a ionizing UV background are invoked, are not
particularly realistic.  Our simulation results agree much better with
the scenario outlined in \citet{Redef_MSP}, which in fact resembles many
of our simulation findings quite closely.

We find that the observed relationship between V-band luminosity and
velocity dispersion is quite well reproduced by our simulations, albeit
with large scatter. The small amount of reliable observational data for
the velocity dispersions leaves it unclear at present whether the larger
scatter we find indicates a problem of the simulations or whether is is
also present in reality. What is comparatively clear though is that the
observed relation between stellar mass and maximum circular velocity is
really tight, a finding that is also reproduced by our simulation
results. On the other hand, the correlation between present-day stellar
mass and star formation rate seen in our simulations seems to be not
nearly as well-defined as in the observational data. This is
related to the fact that we do not find a good correlation between the
present stellar and gaseous masses; many simulated satellites have
comparable stellar masses but differ in their gas fractions by huge
factors. Gas-rich and completely gas-depleted satellites coexist in the
same total and stellar mass regime, rendering a tight correlation with
the star formation rate unlikely.

But perhaps the most significant discrepancy between the simulation
results and observations lies in the inferred mass-to-light ratios.  The
mass-to-light ratios of the simulated galaxies are off by about a factor
of $5$ when compared at face value to the observational estimates. This
means that they are either too massive, or too faint for their mass.
The discrepancy could also be caused by a systematic underestimate of
the total satellite masses in the observations. Due to the difficulty of
reliably determining the `outer edge' of the dark matter halo of an
orbiting satellite, this possibility cannot be easily excluded.

In summary, we find that the current generation of cosmological
hydrodynamic simulations is able to explain many properties of the
observed satellite population surprisingly well. We have shown that
different feedback physics affects the satellite population strongly,
with respect to quantities such as luminosity function, scaling
relations, or star formation histories. This emphasizes the significant
potential of ``near-field cosmology'' within our Local Group to inform
the general theory of galaxy formation. Our work has also shown that it
is not necessarily the physics of cosmic reionization and supernova
feedback alone that is responsible for resolving the missing satellite
problem. In fact, the role of reionization has probably been grossly
overstated in many previous works, while other important feedback, such
as cosmic rays, has been ignored. It will therefore be very interesting
to refine the hydrodynamical simulations further in future work, and to
make them more faithful in capturing all the relevant physics.

\section{Acknowledgements}\label{sec:Acknowledgements}
This research was supported by the DFG cluster of excellence `Origin 
and Structure of the Universe'.

\bibliographystyle{mnras}
\bibliography{bibliography}{}

\end{document}